\documentclass[twocolumn,twocolappendix]{aastex631}
\usepackage{amsmath}

\received{March 4, 2024}
\revised{May 2, 2024}
\accepted{May 3, 2024}

\begin{document}

\title{X-ray Winds In Nearby-to-distant Galaxies (X-WING) - I: Legacy Surveys of Galaxies with Ultrafast Outflows and Warm Absorbers in $z \sim 0$--$4$}

\author[0000-0002-9754-3081]{Satoshi Yamada}
\affiliation{RIKEN Cluster for Pioneering Research, 2-1 Hirosawa, Wako, Saitama 351-0198, Japan; satoshi.yamada@riken.jp}

\author[0000-0002-6808-2052]{Taiki Kawamuro}
\affiliation{RIKEN Cluster for Pioneering Research, 2-1 Hirosawa, Wako, Saitama 351-0198, Japan; satoshi.yamada@riken.jp}

\author[0000-0003-2161-0361]{Misaki Mizumoto}
\affiliation{Science Education Research Unit, University of Teacher Education Fukuoka, 1-1 Akama-bunkyo-machi, Munakata, Fukuoka 811-4192, Japan}

\author[0000-0001-5231-2645]{Claudio Ricci}
\affiliation{Instituto de Estudios Astrof\'isicos, Facultad de Ingenier\'ia y Ciencias, Universidad Diego Portales, Av. Ej\'ercito Libertador 441, Santiago, Chile}
\affiliation{Kavli Institute for Astronomy and Astrophysics, Peking University, Beijing 100871, China}

\author[0000-0002-5701-0811]{Shoji Ogawa}
\affiliation{Institute of Space and Astronautical Science (ISAS), Japan Aerospace Exploration Agency (JAXA), 3-1-1 Yoshinodai, Chuo-ku, Sagamihara, Kanagawa
252-5210, Japan}

\author[0000-0001-6020-517X]{Hirofumi Noda}
\affiliation{Astronomical Institute, Tohoku University, 6-3 Aramakiazaaoba, Aoba-ku, Sendai, Miyagi 980-8578, Japan}

\author[0000-0001-7821-6715]{Yoshihiro Ueda}
\affiliation{Department of Astronomy, Kyoto University, Kitashirakawa-Oiwake-cho, Sakyo-ku, Kyoto 606-8502, Japan}

\author[0000-0003-1244-3100]{Teruaki Enoto}
\affiliation{Division of Physics and Astronomy, Graduate School of Science, Kyoto University, Kitashirakawa, Sakyo-ku, Kyoto 606-8502, Japan}

\author[0000-0001-6402-1415]{Mitsuru Kokubo}
\affiliation{National Astronomical Observatory of Japan, 2-21-1 Osawa, Mitaka, Tokyo 181-8588, Japan}

\author[0000-0002-2933-048X]{Takeo Minezaki}
\affiliation{Institute of Astronomy, School of Science, the University of Tokyo, 2-21-1 Osawa, Mitaka, Tokyo 181-0015, Japan}

\author[0000-0001-6401-723X]{Hiroaki Sameshima}
\affiliation{Institute of Astronomy, School of Science, the University of Tokyo, 2-21-1 Osawa, Mitaka, Tokyo 181-0015, Japan}

\author[0000-0001-5925-3350]{Takashi Horiuchi}
\affiliation{Institute of Astronomy, School of Science, the University of Tokyo, 2-21-1 Osawa, Mitaka, Tokyo 181-0015, Japan}

\author[0000-0002-6068-8949]{Shoichiro Mizukoshi}
\affiliation{Institute of Astronomy, School of Science, the University of Tokyo, 2-21-1 Osawa, Mitaka, Tokyo 181-0015, Japan}



\begin{abstract}
As an inaugural investigation under the X-ray Winds In Nearby-to-distant Galaxies (X-WING) program, we assembled a dataset comprising 132 active galactic nuclei (AGNs) spanning redshifts $z \sim 0$--4 characterized by blueshifted absorption lines indicative of X-ray winds. 
Through an exhaustive review of previous research, we compiled the outflow parameters for 573 X-ray winds, encompassing key attributes such as outflow velocities ($V_{\rm out}$), ionization parameters ($\xi$), and hydrogen column densities.
By leveraging the parameters $V_{\rm out}$ and $\xi$, we systematically categorized the winds into three distinct groups: ultrafast outflows (UFOs), low-ionization parameter (low-IP) UFOs, and warm absorbers. 
Strikingly, a discernible absence of linear correlations in the outflow parameters, coupled with distributions approaching instrumental detection limits, was observed. 
Another notable finding was the identification of a velocity gap around $V_{\rm out} \sim 10,000~{\rm km~s^{-1}}$.
This gap was particularly evident in the winds detected via absorption lines within the $\lesssim$2~keV band, indicating disparate origins for low-IP UFOs and warm absorbers. In cases involving \ion{Fe}{25}/\ion{Fe}{26} lines, where the gap might be attributed to potential confusion between emission/absorption lines and the Fe K-edge, the possibility of UFOs and galactic-scale warm absorbers being disconnected is considered.
An examination of the outflow and dust sublimation radii revealed a distinction: UFOs appear to consist of dust-free material, whereas warm absorbers likely comprise dusty gas. From 2024, the X-Ray Imaging and Spectroscopy Mission (XRISM) is poised to alleviate observational biases, providing insights into the authenticity of the identified gap, a pivotal question in comprehending AGN feedback from UFOs.


\end{abstract}



\section{Introduction} \label{S1-intro}
Outflows stemming from supermassive black holes (SMBHs) are pivotal in not only shaping the active galactic nucleus (AGN) structure \citep[e.g.,][]{Krolik1986, Antonucci1993, Buchner2015, Ramos-Almeida2017, Garcia-Burillo2021, Kawamuro2021} but also influencing the evolution of host galaxies by injecting energy and momentum into the interstellar medium \citep[e.g.,][]{Silk1998, Di-Matteo2005, Fabian2012, Cicone2014, Harrison2017}. Theoretical studies posit that AGN outflows can curb star formation in galaxies \citep[e.g.,][]{Fabian2012, Zubovas2012}, potentially elucidating the observed correlation between the SMBH mass ($M_{\rm BH}$) and velocity dispersion of the host galaxy \citep[e.g.,][]{Kormendy2013}. Owing to the multiphase nature of outflows \citep[e.g.,][]{Laha2021}, comprehending their complete structure and feedback impact across various spatial scales or under different physical conditions remains challenging. Consequently, extensive multi-wavelength studies on multiphase outflows have been conducted \citep[e.g.,][]{Fiore2017, Fluetsch2019, Fluetsch2021, Yamada2021, Yamada2023, Izumi2023}.

Various components of multiphase outflows are discernible across different wavelengths. In X-rays, highly ionized ultrafast outflows (UFOs; \citealt{Tombesi2010d, Gofford2015}) and slower outflows of warm absorbers \citep{McKernan2007, Laha2014b} are observed. In ultraviolet (UV) and optical bands, broad absorption line (BAL) outflows \citep{Mehdipour2023a, Temple2023}, ionized \citep{Fischer2013, Rojas2020, Ruschel-Dutra2021} and neutral outflows \citep{Rupke2005, Rupke2017, Fluetsch2021} are identified. In the near/mid-infrared (IR) band, warm ionized \citep{Muller-Sanchez2011, Bianchin2022}, hot molecular \citep{Davies2014, Riffel2023}, and dusty outflows\footnote{While dust emission displaying no emission/absorption lines makes it challenging to measure its velocity, some studies suggest that polar dust are outflowing by optical spectropolarimetry \citep{Yoshida2011} and IR emission models \citep[e.g.,][]{Honig2017,Stalevski2019,Venanzi2020,Alonso-Herrero2021,Yamada2023}.} \citep{Lopez-Gonzaga2016, Honig2019} are detected. In the far-IR and submillimeter range, warm OH \citep{Stone2016, Gonzalez-Alfonso2017} and cold molecular outflows \citep{Cicone2014, Janssen2016, Lutz2020} are observed. In the radio domain, \ion{H}{1} neutral outflows \citep{Morganti2005, Morganti2016} and relativistic jets \citep{Rafferty2006,Smolcic2017} are identified. Given the fast nature of UFOs ($\sim$0.1--0.3$c$) launched from the proximity of an SMBH ($\sim$10--100$r_{\rm s}$; $r_{\rm s}$ being the Schwarzschild radius), a comparative analysis of UFOs and outer multiphase outflows becomes imperative to comprehend the energetics of multiscale outflows.

\citet{Fiore2017} conducted an investigation of multiphase outflows (UFOs, BALs, optical ionized outflows, and warm/cold molecular outflows) by compiling multiwavelength observations of 94 AGN host galaxies at redshifts $z \sim 0$--6. They identified scaling relationships between outflow properties, AGN characteristics, and host galaxy properties, such as outflow rates, AGN luminosities, and star-formation rates. However, owing to differing outflow types in various samples, the energetic aspects of individual object multiphase outflows remained unclear. Numerous studies \citep[e.g.,][]{Tombesi2015, Mizumoto2019a, Nardini2018, Smith2019, Bonanomi2023} have investigated the relationships between outflow rates and velocities (or distances from the center) of multiphase outflows, revealing substantial dispersions in these parameters. To discern potential physical relationships, constructing a larger sample of AGNs with multiphase outflows and comparing its properties with those of outflows, AGNs, and host galaxies is imperative.

Furthermore, in the realm of X-ray outflows, the relationship between UFOs and warm absorbers remains an open question. The fundamental properties of these outflows are dictated by their velocities ($V_{\rm out}$) and ionization parameters ($\xi$) and the hydrogen column densities of the ionized gas ($N_{\rm H}$). The ionization parameter is defined as $\xi = L_{\rm ion}/n_{\rm H}r^2$, where $L_{\rm ion}$ represents ionizing luminosity in the 13.6 eV to 13.6 keV band or the 1--1000 Ryd range, $n_{\rm H}$ is the hydrogen number density, and $r$ is the distance from the source. Some theoretical models predict a positive correlation between $V_{\rm out}$ and $\xi$ for X-ray winds (i.e., UFOs and outflowing warm absorbers) in scenarios such as the radiative-driven wind model \citep[see e.g., ][]{King2003, King2010} and magnetohydrodynamically (MHD) driven model \citep[e.g.,][]{Blandford1982, Fukumura2010, Fukumura2017}. While prior X-ray observations of approximately 50 X-ray winds in several tens of AGNs \citep[e.g.,][]{Tombesi2013a, Laha2014b} support a positive correlation, elucidating the impact of observational biases and direct connection between UFOs and warm absorbers remains challenging. Thus, expanding the sample size of X-ray winds is crucial for mitigating observational biases and comprehensively understanding their structure.

To amass the largest database of UFOs and larger-scale multiphase outflows, we initiated the X-ray Winds In Nearby-to-distant Galaxies (X-WING) project. The X-WING sample comprises AGNs exhibiting X-ray winds (UFOs and warm absorbers). It encompasses AGNs with warm absorbers where UFOs have not been detected, as they can also be potential candidates for AGNs with UFOs, some of which may be detectable via high-resolution X-ray spectroscopy with the X-Ray Imaging and Spectroscopy Mission (XRISM) launched in September 2023. We intend to present a comprehensive database of AGN multiphase outflows, along with host galaxy properties, across three papers. This database will facilitate investigations into the structures, driving mechanisms, and energetics of X-ray winds and other multiphase outflows.

In this study (Paper I), we meticulously examined previous studies on X-ray winds, presenting the most extensive database of outflow parameters for 573 X-ray winds in 132 AGNs to date and discussing their characteristics. The subsequent sections are organized as follows. Section~\ref{S2-database} details X-WING and X-ray wind databases. Section~\ref{S3-results} reports X-ray wind properties ($V_{\rm out}$, $\xi$, and $N_{\rm H}$) and potential biases in our results (Sections~\ref{sub3-1-winds}--\ref{sub3-3-Fe}). Section~\ref{sub3-4-location} explores the location and driving mechanism of X-ray winds, while Section~\ref{sub3-5-xrism} discusses anticipated new findings using XRISM. Finally, Section~\ref{S4-summary} summarizes the main conclusions. Paper II will present UV-to-radio spectral energy distribution (SED) fitting results and compare AGN and host activities with X-ray wind outflow rates. Paper III will present a catalog of multiphase outflow rates and discuss the energetics of all outflows (S. Yamada et al., in preparation). Throughout this study, ${\log}V_{\rm out}$, ${\log}\xi$, and ${\log}N_{\rm H}$ represent the logarithms of $V_{\rm out}$, $\xi$, and $N_{\rm H}$ in km~s$^{-1}$, erg~s$^{-1}$~cm, and cm$^{-2}$, respectively. We adopt the cosmological parameters $H_{\rm 0} = 70$~km~s$^{-1}$ Mpc$^{-1}$, $\Omega_{\rm M} = 0.3$, and $\Omega_{\rm \Lambda} = 0.7$.
\\

\section{Sample and X-ray Wind Catalog} \label{S2-database}
\subsection{X-WING AGNs} \label{sub2-1-sample}
We attempted to identify all galaxies wherein X-ray winds (UFOs and outflowing warm absorbers) were reported by the end of 2023, as detected by their X-ray blueshifted ionized absorption lines.
In this study, we focused on the X-ray instruments utilized to identify X-ray winds over the past 20 years.
(1) Chandra/High Energy Transmission Grating Spectrometer (HETGS), 
(2) Chandra/Low Energy Transmission Grating Spectrometer (LETGS),
(3) Chandra/Advanced CCD Imaging Spectrometer (ACIS), 
(4) XMM-Newton/Reflection Grating Spectrometer (RGS),
(5) XMM-Newton/European Photon Imaging Camera (EPIC) PN, 
(6) XMM-Newton/EPIC MOS,
(7) Nuclear Spectroscopic Telescope Array (NuSTAR)/Focal Plane Modules (FPM),
(8) Suzaku/X-ray Imaging Spectrometer (XIS), and
(9) Swift/X-Ray Telescope (XRT).
The description and performance of these instruments are summarized in Appendix~\ref{AppendixA1-inst} and Table~\ref{TA1-inst}.
Based on studies using these instruments, we first referred to the following papers.

\begin{figure}
    \centering
    \includegraphics[keepaspectratio, scale=0.11]{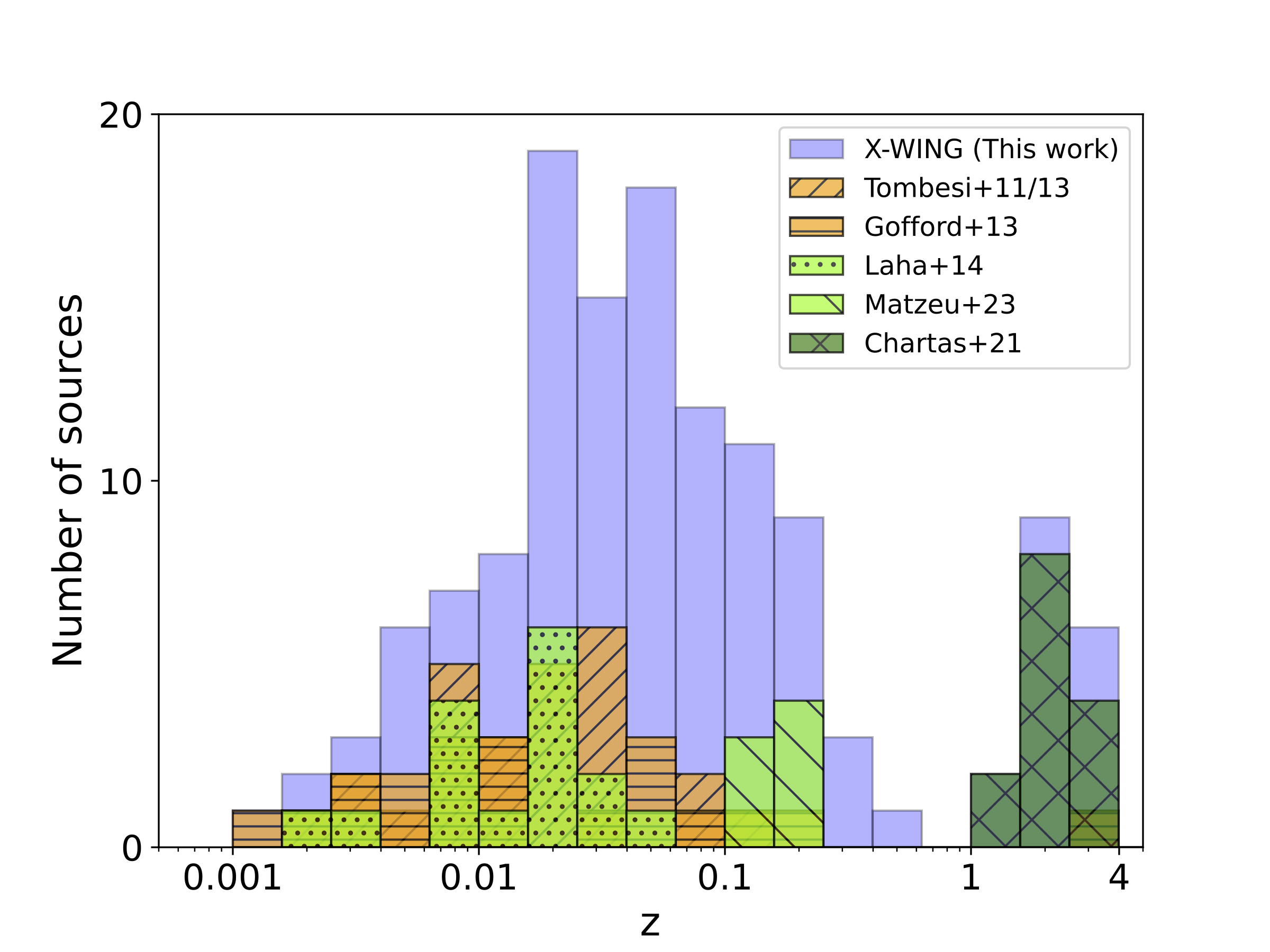}
    \caption{Histogram of redshift for the whole X-WING sample, compared with the sample of \citet{Tombesi2011a,Tombesi2013a}, \citet{Gofford2013}, \citet{Laha2014b}, \citet{Matzeu2023}, and \citet{Chartas2021}.
    \\}
\label{F1-z-hist}
\end{figure}

\begin{enumerate}
    \item \citet{Tombesi2010d,Tombesi2011a,Tombesi2012a,Tombesi2013a}: UFOs detected with XMM-Newton/PN and/or outflowing warm absorbers reported in the prior studies with Chandra/HETGS, LETGS, XMM-Newton/RGS, PN, and MOS, in 26 out of 42 local radio-quiet AGNs;

    \item \citet{Gofford2013,Gofford2015}: UFOs detected with Suzaku/XIS in 20 out of 51 local AGNs;

    \item \citet{Laha2014b,Laha2016b}: outflowing warm absorbers detected with XMM-Newton/RGS and PN in 16 out of 26 local AGNs;

    \item \citet{Matzeu2023}: UFOs in the $0.1 \lesssim z \lesssim 0.4$ galaxies detected with XMM-Newton/PN and MOS in 7 out of 22 sources, as part of the SUpermassive Black hole Winds in the x-rAYS (SUBWAYS) program;

    \item \citet{Chartas2021} reported the detection of UFOs using Chandra/ACIS and XMM-Newton/PN and MOS in a sample of 14 Quasi-Stellar Objects (QSOs), with 12 of them being lensed QSOs. These objects span a redshift range of 1.41--3.91.
    
\end{enumerate}
In these studies, the significance thresholds of UFO detections utilized are typically at $\gtrsim$3$\sigma$ level, except for some sources ($\sim$2$\sigma$, see Section~\ref{sub2-2-winds}).
In addition to the works, various systematic surveys \citep[e.g.,][]{McKernan2007, Patrick2012, Tombesi2014a, Mehdipour2019} have contributed to our understanding of the field.
Following this, we conducted a comprehensive review of previous investigations into X-ray winds using data from Chandra, XMM-Newton, NuSTAR, Suzaku, and Swift, spanning the period from approximately 2000 to 2023. From a compilation of 285 papers, we identified 132 galaxies within the redshift range of $z \sim 0$--4, as detailed in Appendix~\ref{AppendixB-database} and Table~\ref{TB2-sample}.
For a subset of six out of the 132 AGNs, blueshifted absorption lines were not discerned through spectral fitting but were detected using spectral variability, as reported by \citet{Igo2020}. This study estimated the outflow velocities ($V_{\rm out}$) for several tens of targets, and we referenced their findings solely for the six AGNs categorized as X-WING candidates in Table~\ref{TB2-sample}.
To provide context, Figure~\ref{F1-z-hist} compares the redshift distribution of X-WING AGNs with those from prior studies. Notably, our sample size is 4--10 times larger than those in previous systematic surveys, owing to extensive follow-up observations of X-ray winds over the past 20 years.

In addition, we present an overview of the fundamental properties of these designated targets. The redshifts and luminosity distances ($D_{\rm L}$) were predominantly obtained from the CDS Portal (derived from SIMBAD)\footnote{\url{http://cdsportal.u-strasbg.fr/}}. For nearby objects within a distance of $<$50~Mpc, we referenced the most recent redshift-independent distances as determined by \citet{Koss2022b}.

\begin{figure}
    \centering
    \includegraphics[keepaspectratio, scale=0.125]{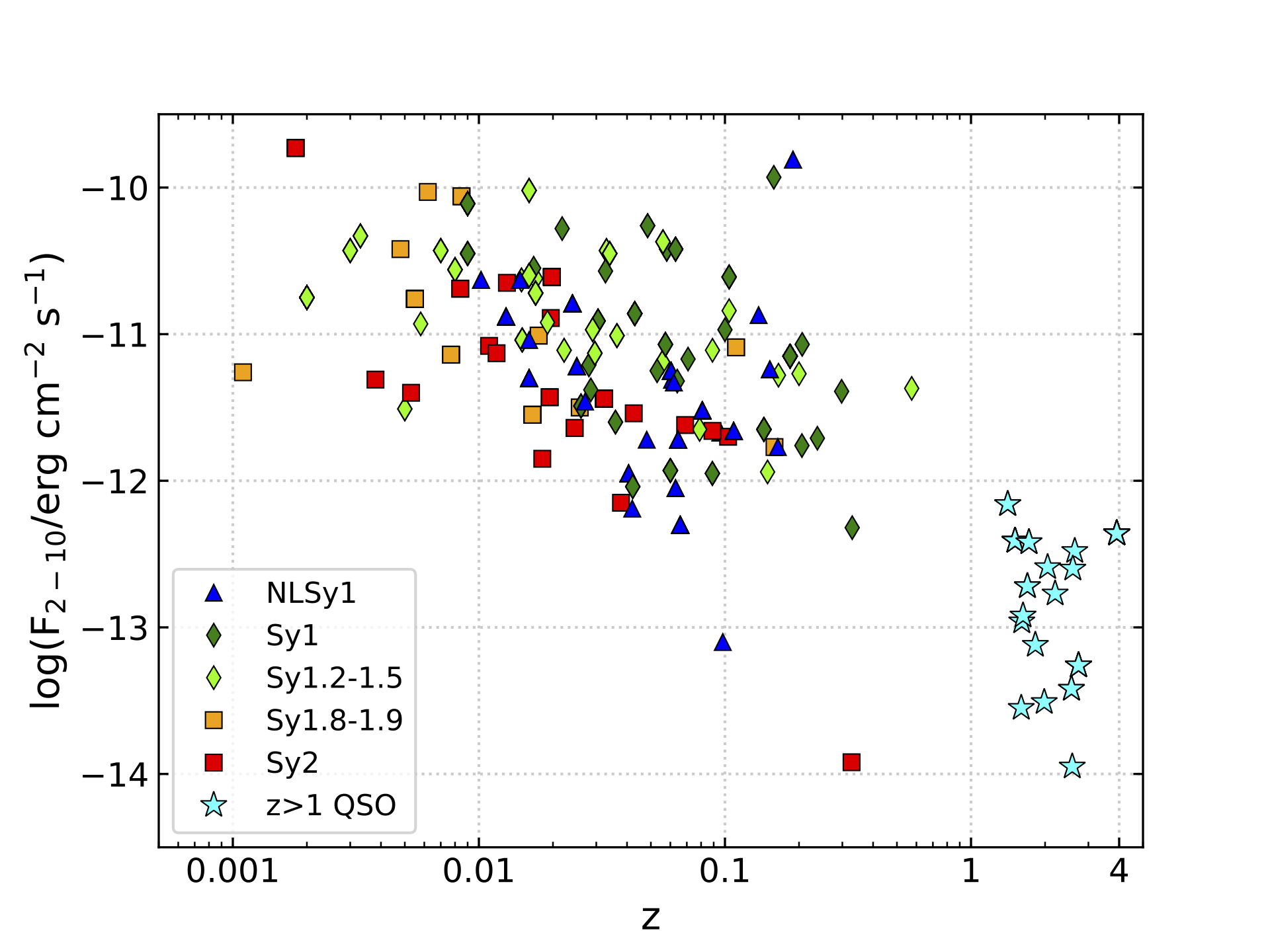}
    \includegraphics[keepaspectratio, scale=0.125]{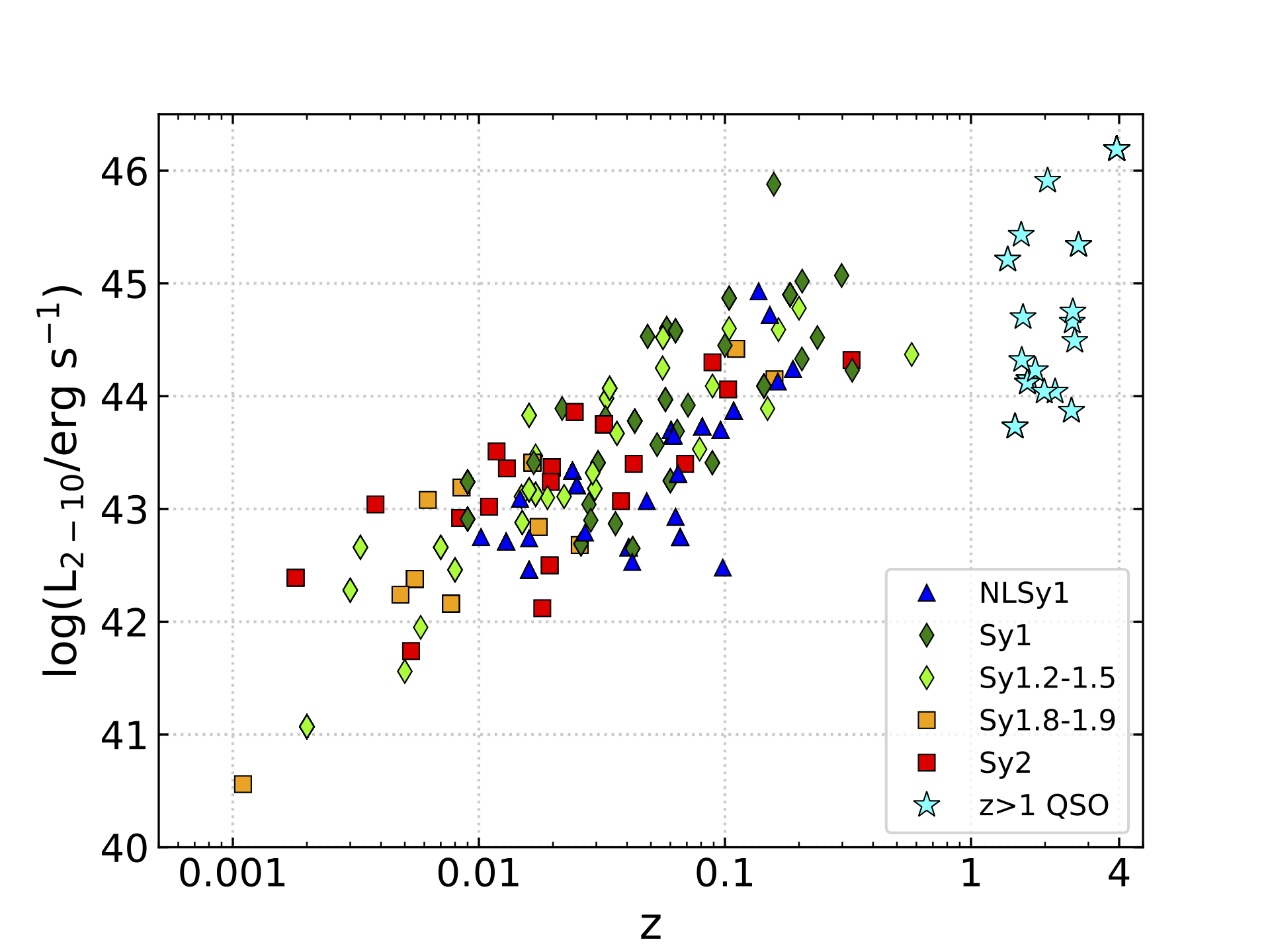}
    \caption{(Top) logarithmic observed 2--10~keV flux ($F_{\rm 2-10}$) vs. $z$.
    (Bottom) logarithmic intrinsic AGN luminosity in the 2--10~keV band ($L_{\rm 2-10}$) vs. $z$.
    Symbols mark optical classification of NLSy1 (blue triangle), Seyfert 1.0 (green diamond), 1.2--1.5 (light-green diamond), 1.8--1.9 (orange square), 2.0 (red square), and $z > 1$ QSOs (cyan star).
    \\}
\label{F2-z-Lx}
\end{figure}

To categorize their optical AGN types, we employed a classification system distinguishing between narrow-line Seyfert 1 (NLSy1), Seyfert 1.0, 1.2, 1.5, 1.8, 1.9, and 2.0 for local AGNs, and high-$z$ QSOs for AGNs at $z > 1$ \citep[e.g.,][]{Veron-Cetty2010}. The determination of the black hole masses ($M_{\rm BH}$) was primarily grounded in reverberation mapping studies \citep[e.g.,][]{Woo2002, Bentz2015, Du2015, Bentz2023a, Woo2024}, assuming a standard virial factor of 4.3 \citep{Grier2013} for measurements relying on the velocity dispersions of broad emission lines (e.g., H$\beta$). In instances where $M_{\rm BH}$ was unspecified, we consulted the values adopted in the BAT AGN Spectroscopic Survey (BASS) project \citep{Koss2022b, Koss2022c} or references focusing on X-ray winds.

We have compiled the observed 2--10~keV fluxes ($F_{\rm 2-10}$) and intrinsic (deabsorbed) 2--10~keV luminosities ($L_{\rm 2-10}$) for our designated targets, primarily from the BASS catalog \citep{Ricci2017d} and XMM-Newton works \citep[e.g.,][]{Bianchi2009a}, as the typical or time-averaged fluxes. It is noteworthy that approximately 10\% of our targets were classified as changing-look AGNs, demonstrating substantial alterations in optical types and/or significant variations in line-of-sight $N_{\rm H}$ (\citealt[e.g.,][]{LyuBing2022}; also refer to \citealt{Noda2018, Ricci2023c}). Detailed discussions on these changing-look AGNs will be provided in X-WING Paper II.

For QSOs with redshifts exceeding 1, the 2--10~keV flux ($F_{\rm 2-10}$) was not adjusted for lensing magnification. In contrast, corrected values were applied for $L_{\rm 2-10}$ and $L_{\rm ion}$, considering the magnification factors stipulated in \citet{Chartas2007b, Chartas2021}. All luminosities cited in the references were transformed into values under the assumption that $D_{\rm L}$ was employed.
We also examined the reported number of X-ray winds (refer to Section~\ref{sub2-2-winds}) and investigated the presence or absence of UFOs (where $V_{\rm out} \geqslant 10,000$~km~s$^{-1}$; see Section~\ref{sub3-1-winds}). In conclusion, these findings have been succinctly summarized in Table~\ref{TB2-sample}.

To date, X-WING is the largest catalog of AGNs with X-ray winds (see Figure~\ref{F1-z-hist}).
Figure~\ref{F2-z-Lx} shows the ${\log}F_{\rm 2-10}$--$z$ and ${\log}L_{\rm 2-10}$--$z$ plots.
The X-WING AGNs cover a wide range of $F_{\rm 2-10}$ ($\sim$10$^{-14}$--10$^{-9.5}$~erg cm$^{-2}$~s$^{-1}$), $L_{\rm 2-10}$ ($\sim$10$^{40}$--10$^{46}$~erg s$^{-1}$), and redshifts ($\sim$0--4).
Therefore, the X-WING database reduces detection biases, helping us to understand the general properties of X-ray winds.

\begin{figure*}
    \centering
    \includegraphics[keepaspectratio, scale=0.22]{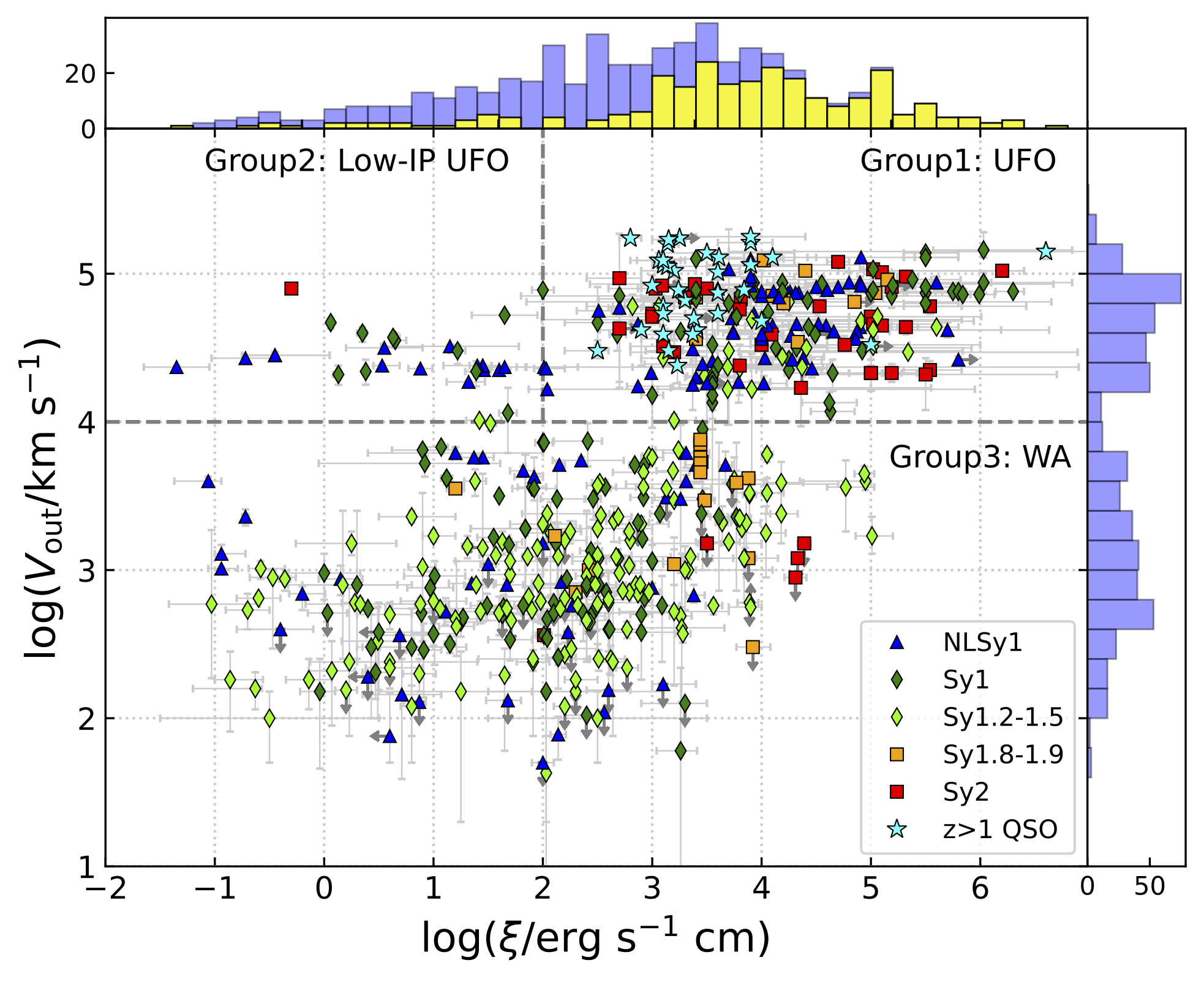}
    \caption{Logarithmic $V_{\rm out}$ vs. logarithmic $\xi$.
    Symbols indicate the optical classification.
    Dashed lines show the thresholds of the three groups of X-ray winds: UFO, low-IP UFO, and warm absorber (WA).
    The top and right histograms indicate the distributions of the parameters.
    The yellow histogram represents the UFOs and low-IP UFOs (i.e., with ${\log}V_{\rm out} \geqslant 4$).
    \\}
\label{F3-outflow-opt-1}
\end{figure*}

\subsection{X-ray Winds} \label{sub2-2-winds}
In Table~\ref{TB3-outflow}, we have presented data on 573 X-ray winds, including their outflow parameters: $V_{\rm out}$, $\xi$, and $N_{\rm H}$ (refer to Section~\ref{S1-intro}). We have also endeavored to compile $L_{\rm ion}$ values.
The determination of photoionization parameters in previous studies followed the expression $U = Q/(4 \pi r^2 c n_e)$, where $Q$ represents the ionizing photon rate, $r$ is the distance from the source, and $n_e$ is the electron density \citep{Netzer1987, Netzer1996}. For 40 out of the 573 X-ray winds, we employed an assumed conversion between $\log U$ and $\log \xi$ using the relation ${\log}U = {\log}\xi - 1.75$, based on a typical AGN SED \citep[e.g.,][]{Proga2004}.
Additionally, Table~\ref{TB3-outflow} provides information on the start dates of observations, the instruments employed for line detection, outflow types of the X-ray winds (discussed in Section~\ref{sub3-1-winds}), the chosen X-ray wind model (refer to Appendix~\ref{AppendixA2-models}), and the methodology for line detection, whether with or without the \ion{Fe}{25} He$\alpha$ (6.697~keV)/\ion{Fe}{26} Ly$\alpha$ (6.966~keV) lines (refer to Section~\ref{sub3-3-Fe}).
For all X-ray winds, individual OBSIDs underwent scrutiny, and in cases where outflows were reported redundantly, priority was given to the most recent or detailed analyses, which often involved broadband spectra and the latest X-ray models.
Finally, we have compiled a summary of the OBSIDs, including references and pertinent papers on duplicated reports, in Table~\ref{TB4-obsid}.

Verification was conducted to ensure that nearly all reported winds were accounted for in the 285 references by identifying the presence of blueshifted absorption lines. However, assessing the significance of these absorption lines proved challenging due to variations in the methodology employed for line detection across the references (refer to Appendix~\ref{AppendixA-inst}).
In this investigation, we meticulously compiled data on all potentially existing X-ray winds within our catalog. 
Despite its limited occurrence, the catalog encompasses values associated with $\sim$2$\sigma$ line detection \citep[as exemplified by][]{Lanzuisi2016} and $V_{\rm out}$ for the six X-WING candidates (outlined in Section~\ref{sub2-1-sample}).
For most of the X-WING candidates, only $V_{\rm out}$ were constrained but not $\xi$ and $N_{\rm H}$.

We also presented additional information on the uncertain wind candidates selected by three primary criteria.
Initially, we classified Compton-thick (CT) AGNs with a column density of neutral absorbers exceeding $10^{24}$~cm$^{-2}$ as highly ambiguous candidates. 
This classification arises due to the potential artifacts caused by strong Fe K$\alpha$ line and Fe-K edge features. 
Using the results from broad X-ray spectral fittings \citep[e.g.,][]{Ricci2017d, Yamada2021}, we identified two CT AGNs, NGC~1068 and NGC~6240 in our sample.
These AGNs exhibit complex X-ray spectra due to the multi-layer absorption (NGC~1068; \citealt{Bauer2015}) and mixed emission from two AGNs (NGC 6240; \citealt{Puccetti2016}).
Secondly, we scrutinized AGNs whose X-ray wind features were low significance (i.e., $\sim$2$\sigma$ levels), PG~1202+281 \citep{Matzeu2023} and PG~1247+268 \citep{Lanzuisi2016}.
The significance of the absorption line in PG~0844+349 is also controversial \citep{Pounds2003d,Brinkmann2006}.
Finally, we investigated cases with no explicit discussion of the detection significance.
Considering the look-elsewhere effect, it would be challenging to indicate how much these winds are reliable.
\textcolor{black}{These sources should be considered as AGNs with uncertain X-ray winds}, denoted by ``(*)'' in column (10) of Table~\ref{TB2-sample}.
Notably, \textcolor{black}{these AGNs (18) and their X-ray wind detections (24) represent a small fraction (13.6\% and 4.2\%) of the entire sample.}
\textcolor{black}{To reduce the impact of the look-elsewhere effect on the wind detection significance} as much as possible, we primarily referred to the results for the entire sample based on the X-ray spectral fitting with the ionized absorber models (e.g., \textsc{xstar} and \textsc{spex}; see Appendix~\ref{AppendixA2-models}), which can detect the multiple absorption line features.
For a more quantitative evaluation, we plan to reanalyze all the archival data in the entire sample by a uniformed method with a simple Gaussian model, \textsc{xstar}, and \textsc{spex} (S. Yamada et al., in preparation).
\textcolor{black}{The main conclusions of this study are nearly unchanged regardless of the presence of a small fraction of low-significance winds, and we thus utilize the whole sample in the following analysis.}

Since we referred to the wind parameters from the various previous works, this study does not present a secured wind catalog but provides the largest database of X-ray wind candidates to date.
The primary strength of this catalog lies in its ability to provide a more comprehensive representation of the general characteristics of X-ray winds, which have been extensively surveyed over time with X-ray instruments, in comparison to previous works. To enhance clarity, whenever literature presented 1$\sigma$ errors, we standardized them to values at the 90\% confidence level by applying a multiplication factor of 1.645. Consequently, all errors and upper/lower limits in this study are consistently presented at the 90\% confidence level.

\begin{figure*}
    \centering
    \includegraphics[keepaspectratio, scale=0.12]{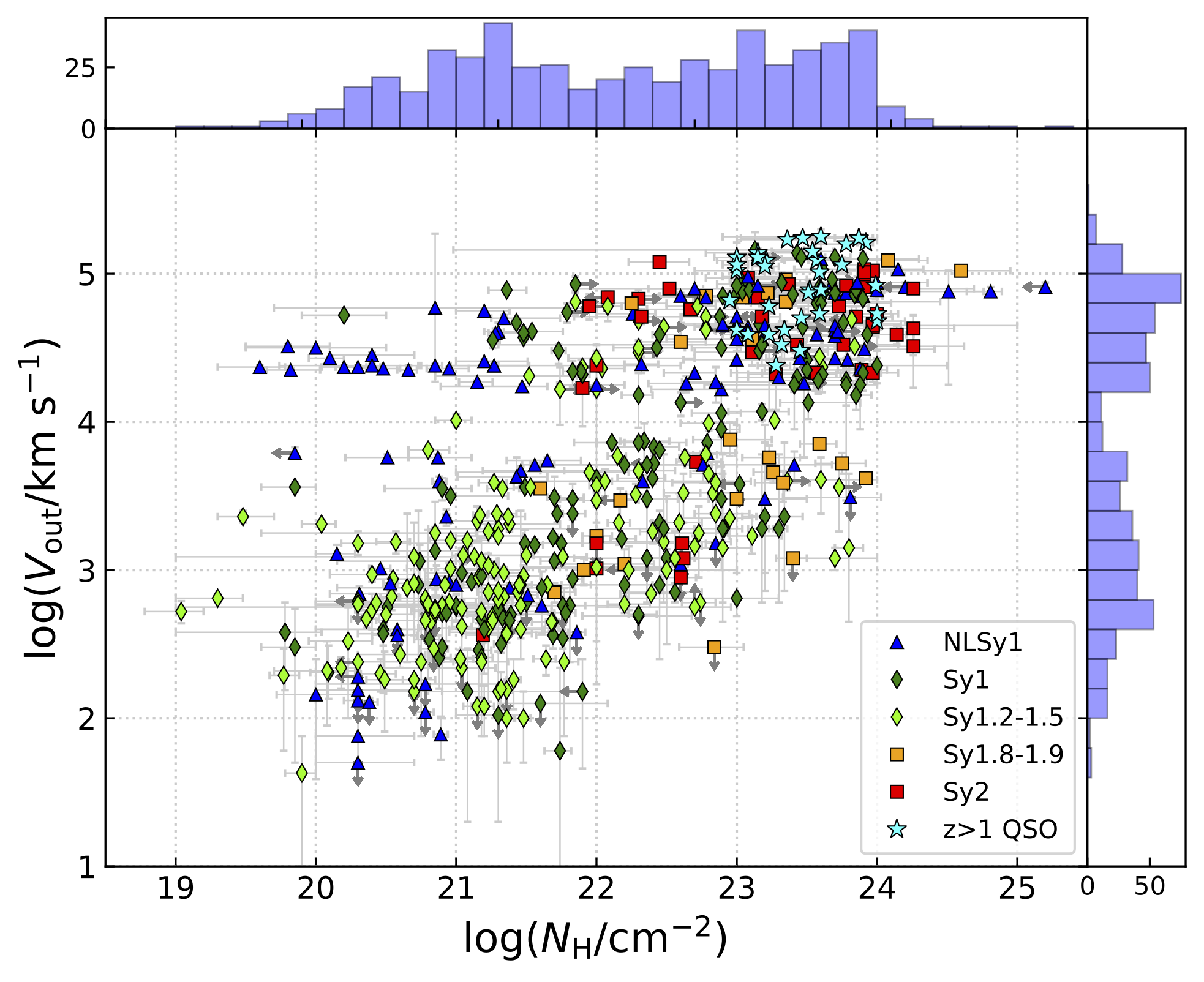}
    \includegraphics[keepaspectratio, scale=0.12]{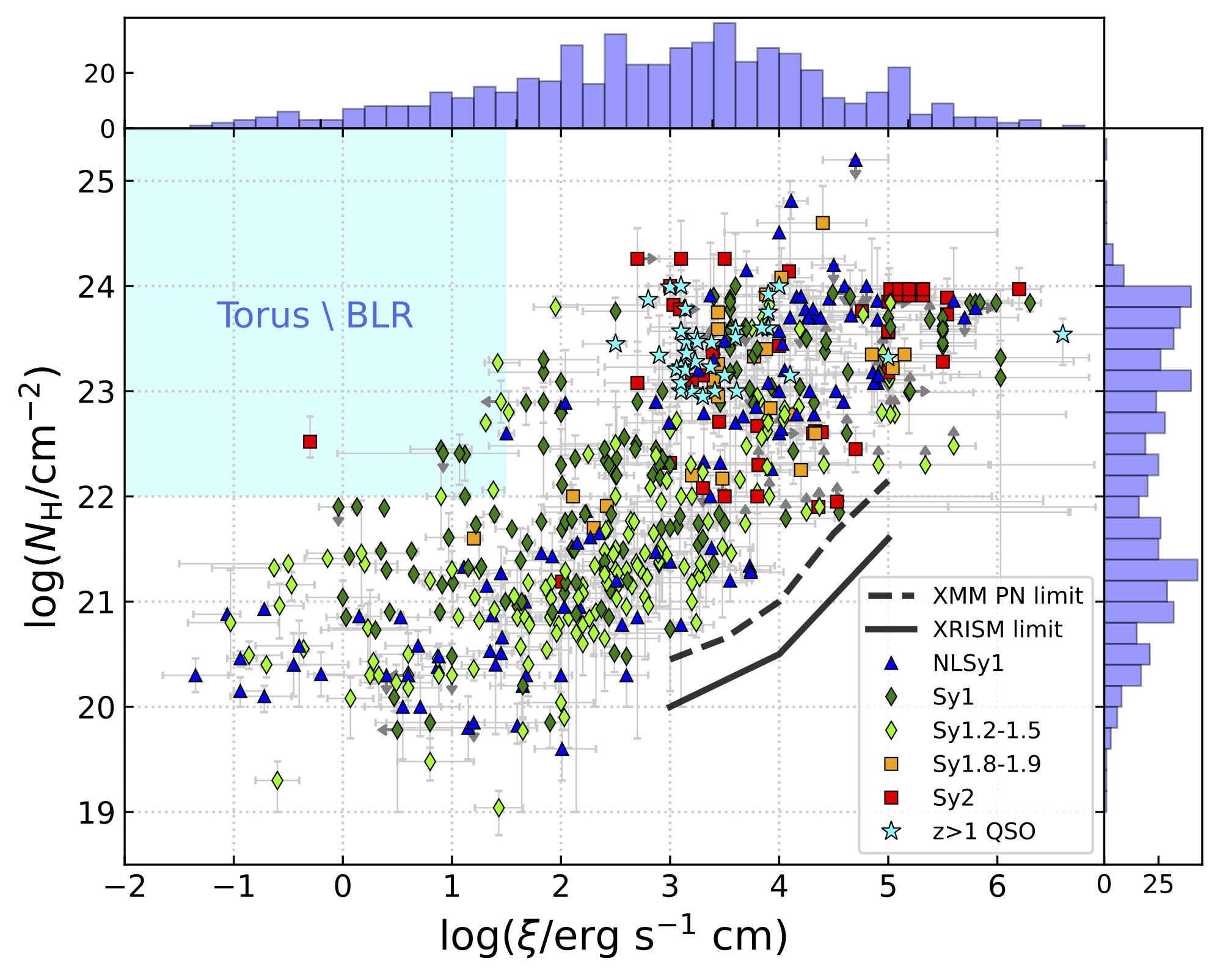}
    \includegraphics[keepaspectratio, scale=0.12]{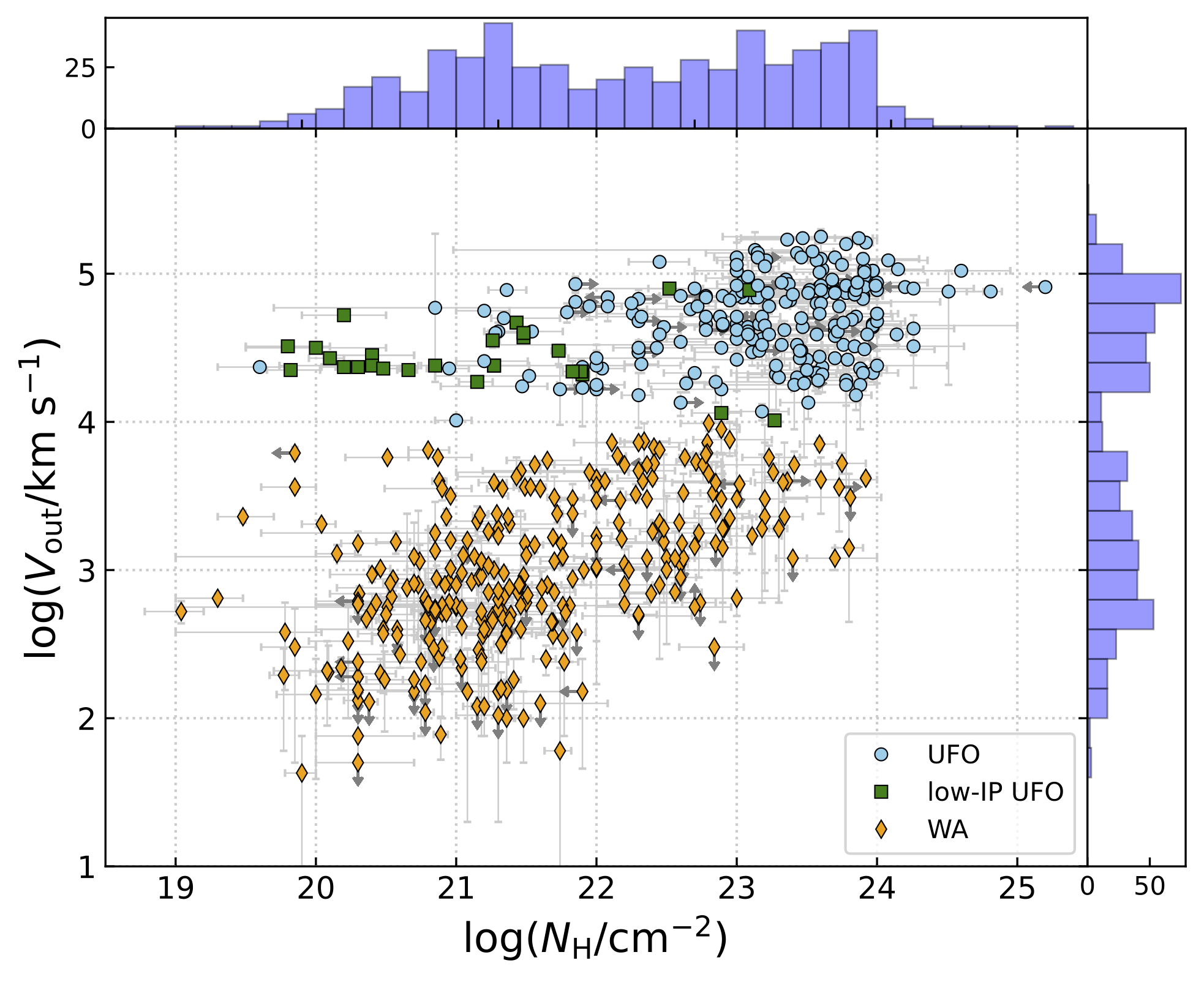}
    \includegraphics[keepaspectratio, scale=0.12]{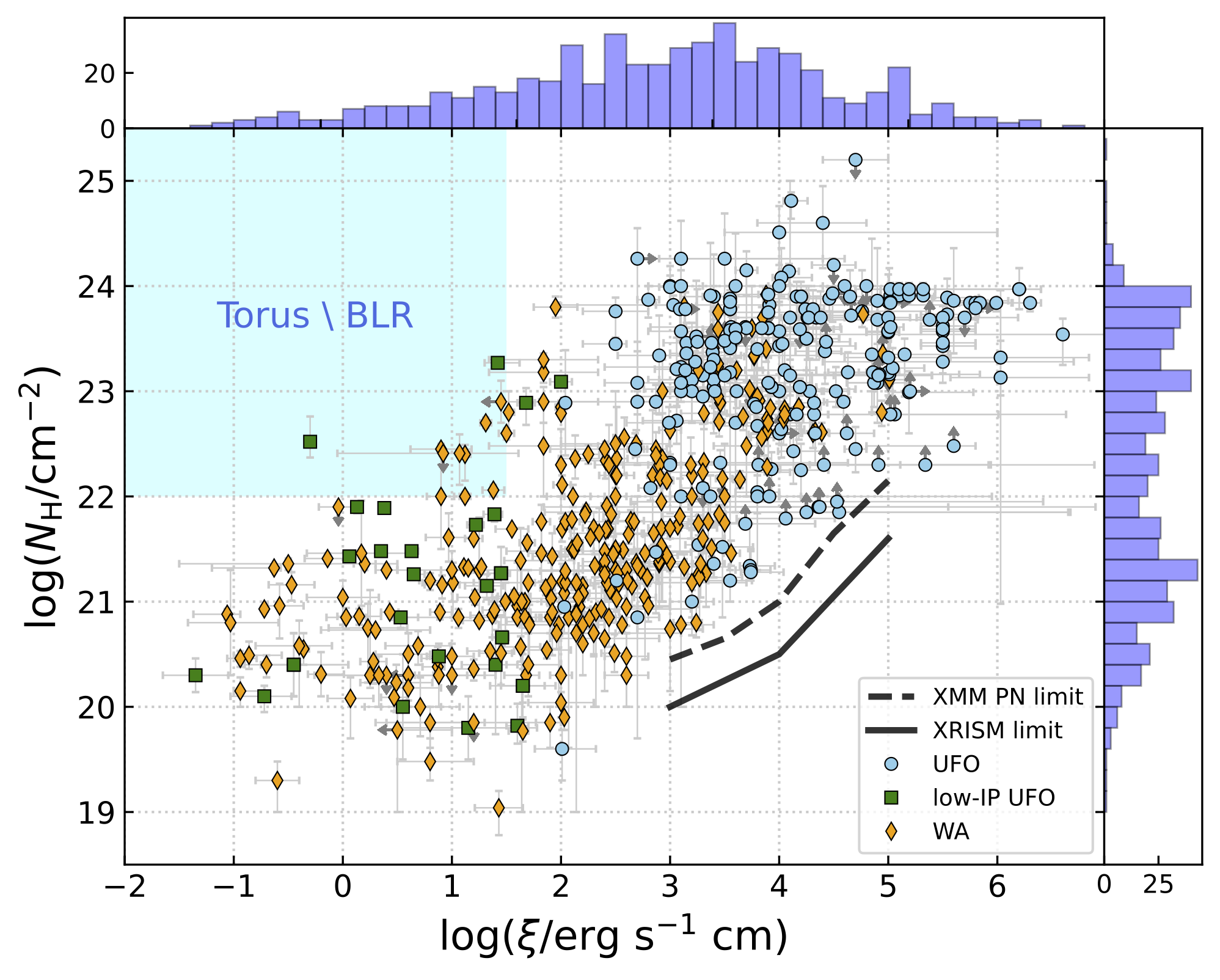}
    \caption{(Top-left) log$V_{\rm out}$ vs. log$\xi$.
    (Top right) log$N_{\rm H}$ versus log$\xi$.
    Symbols indicate the optical classification.
    The cyan region shows typical values for BLRs or tori.
    (Bottom-left) log$V_{\rm out}$ vs. log$\xi$.
    (Bottom right) Log $N_{\rm H}$ versus log$\xi$.
    Symbols mark the outflow groups: UFO, low-IP UFO, and WA, as classified in Figure~\ref{F3-outflow-opt-1}.
    \\}
\label{F4-outflow-opt-2}
\end{figure*}

\section{Results and Discussion} \label{S3-results}
\subsection{X-ray Wind Properties} \label{sub3-1-winds}
To comprehend the characteristics of X-ray winds in the X-WING catalog, we have generated three comparison plots illustrating the relationships between the three outflow parameters ($V_{\rm out}$, $\xi$, and $N_{\rm H}$) in Figures~\ref{F3-outflow-opt-1} and~\ref{F4-outflow-opt-2}. The data points representing the winds are categorized based on optical AGN types: narrow-line Seyfert 1 (NLSy1), Seyfert 1.0, 1.2--1.5, 1.8--1.9, 2.0, and $z > 1$ type-1 QSOs.
Notably, Seyfert 1.8--2.0 galaxies exhibit large values at $V_{\rm out}$, $\xi$, and $N_{\rm H}$.
Conversely, NLSy1 and Seyfert 1.0--1.5 demonstrate broader distributions across these parameters. It is worth considering that many Seyfert 1.8--2.0 galaxies host obscured AGNs with high inclination angles, making it challenging to detect their warm absorbers with low $\xi$ and $N_{\rm H}$ in the soft X-ray band. This difficulty arises because soft X-ray fluxes are often heavily obscured by neutral absorbers in the broad-line regions (BLRs) and dusty tori.
A similar detection bias may be introduced for high-$z$ QSOs, as the rest-frame soft X-rays become invisible due to redshift. Despite the potential bias introduced by these sources, \textcolor{black}{we have verified that even excluding these sources from the following analysis} (Seyfert 1.8--2.0 galaxies and high-$z$ QSOs), the primary outcomes in this study remain unaltered.

\textcolor{black}{High-luminosity AGNs with $L_{\rm 2-10} > 10^{44}$~erg~s$^{-1}$ at $z < 1$ have a similar distribution in wind parameters to the high-$z$ QSOs, with the advantage of also having low-velocity measurements, which are precluded to high-$z$ QSOs.
We found a possibility that they have a higher fraction of UFOs relative to the low-velocity winds compared with lower-luminosity AGNs.
We will present proper bolometric AGN luminosities ($L_{\rm bol}$) by the multiwavelength SED analysis and discuss the potential dependence of the $L_{\rm 2-10}$, $L_{\rm bol}$, and Eddington ratio ($\lambda_{\rm Edd}$) on the wind parameters in Paper II.}

In the ${\log}V_{\rm out}$--${\log}\xi$ plot (Figure~\ref{F3-outflow-opt-1}), we found a double-peaked distribution of $V_{\rm out}$ (see also Section~\ref{sub3-3-Fe}).
For the distribution of ${\log}(V_{\rm out}/{\rm km~s^{-1}}) > 4$ (yellow histogram), most winds have ${\log}(\xi/{\rm erg~s^{-1}~cm}) \geqslant 2$.
For convenience, we categorized X-ray winds into three groups:
\begin{enumerate}
    \item[(1)] UFOs, with ${\log}V_{\rm out} \geqslant 4$ and ${\log}\xi \geqslant 2$;

    \item[(2)] low-ionization-parameter (low-$\xi$ or low-IP) UFOs, with ${\log}V_{\rm out} \geqslant 4$ but ${\log}\xi < 2$;

    \item[(3)] and warm absorbers, with ${\log}V_{\rm out} < 4$.
\end{enumerate}
While the term ``warm'' absorbers conventionally denote ionized gas at a temperature of approximately ${\sim}10^5$~K higher than cold neutral gas yet considerably cooler than the thermal plasma \citep[e.g.,][]{Reynolds1995}, our study extends this definition to encompass absorbers at both higher and lower temperatures, characterized solely by their modest $V_{\rm out}$.
In a seminal work, \citet{Tombesi2013a} highlighted a potential correlation between ${\log}V_{\rm out}$ and ${\log}\xi$ within X-ray winds (UFOs and warm absorbers) across 35 Seyfert 1 galaxies. Subsequently, \citet{Laha2014b} conducted a follow-up investigation focusing on warm absorbers in 26 Seyfert galaxies. Their findings revealed a shallow slope in the ${\log}V_{\rm out}$--${\log}\xi$ linear regression, deviating from the distribution observed for UFOs. Both studies acknowledged the potential influence of observational biases stemming from instrumental performance, such as energy resolution and detection sensitivities.
Recent research has unveiled the existence of low-IP UFOs \citep[as seen in, e.g.,][]{Gupta2013c,Serafinelli2019}, introducing complexity to their distribution. Even with the exclusion of low-IP UFOs, our outcomes reveal a substantial dispersion of X-ray winds within the X-WING AGNs.
Similarly, the distribution in the ${\log}V_{\rm out}$--${\log}N_{\rm H}$ plane (left panels of Figure~\ref{F4-outflow-opt-2}) does not imply a straightforward linear correlation, even upon excluding low-IP UFOs exhibiting ${\log}V_{\rm out} > 4$ and ${\log}(N_{\rm H}/{\rm cm^{-2}}) \lesssim 22$ (also evident in the right-bottom panel).

Certain investigations posit that low-IP UFOs may originate in proximity to the SMBH and extend to a larger scale of approximately $\sim$100~pc (e.g., \citealt{Serafinelli2019}). Notably, low-IP UFOs exhibit small ionization parameters ($\xi$) (proportional to $r^{-2}$ or indicative of substantial distances) and hydrogen column densities ($N_{\rm H}$) even at high velocities exceeding 10,000~km~s$^{-1}$.
The low-IP UFOs are detected in NLSy1 and Seyfert 1 galaxies, except for 1ES~1927+654, a Seyfert 2 galaxy exhibiting little X-ray obscuration \citep{Gallo2013b}.

In contrast to the ${\log}V_{\rm out}$--${\log}\xi$ and ${\log}V_{\rm out}$--${\log}N_{\rm H}$ plots, the distribution of ${\log}\xi$--${\log}N_{\rm H}$ (depicted in the right panels of Figure~\ref{F4-outflow-opt-2}) exhibits a potential linear correlation. Nonetheless, the dispersion of $N_{\rm H}$ within each $\xi$ bin (approximately 3--4~dex) surpasses that observed in prior studies (around 2~dex; e.g., \citealt{Tombesi2013a}). This variance likely hinges on observational biases.
Here, we examine two biased regions characterized by low-$\xi$ (${\lesssim}10^3$~erg~s$^{-1}$~cm) and high-$\xi$ (${\gtrsim}10^3$~erg~s$^{-1}$~cm) lines, detectable in the $\lesssim$2~keV and $\sim$6~keV bands, respectively \citep[see e.g.,][]{Gallo2023}. The cyan-shaded region, indicative of low $\xi$ and high $N_{\rm H}$, is dominated by clouds optically thicker than typical X-ray winds, essentially comprising BLRs or tori. Consequently, absorption line features are scarce due to the obscuration of the soft continuum in the presence of such clouds in the line of sight (e.g., Seyfert 2). For these obscured AGNs, challenges arise in identifying X-ray winds due to the contribution of optically thin thermal emission from the plasma in the host galaxy \citep{Smith2001} and AGN-scattered emission \citep[e.g.,][]{Kawamuro2016,Gupta2021}.
Conversely, in the vacant area, the lines corresponding to high-$\xi$ and low-$N_{\rm H}$ are weak in the $\gtrsim$6~keV band. This weakness stems from the low $N_{\rm H}$ of ionized gas, resulting in feeble line features, potentially observed as ``bare'' AGNs \citep[e.g.,][]{Walton2013,Nandi_Prantik2023,Porquet2024}. A recent X-ray simulation of the radiation-hydrodynamic AGN model \citep{Wada2016,Ogawa2022} also indicates that edge-on (low-$\xi$/high-$N_{\rm H}$) and face-on (high-$\xi$/low-$N_{\rm H}$) AGNs display no visible absorption lines, whereas AGNs with moderate inclination angles exhibit absorption features.
In our sample, where X-ray winds were predominantly detected in AGNs with moderate inclination angles, no distinct variations were noted in the distributions among optical types (Seyfert 1--2). This implies that the dependence of AGNs on inclination angle is not stringent. In essence, the ${\log}\xi$--${\log}N_{\rm H}$ distribution primarily underscores rough disparities between edge-on, face-on, and other AGNs, with the breadth of the distribution of detected X-ray winds being determined by detection sensitivity (see Section~\ref{sub3-2-bias}).

\begin{figure*}
    \centering
    \includegraphics[keepaspectratio, scale=0.18]{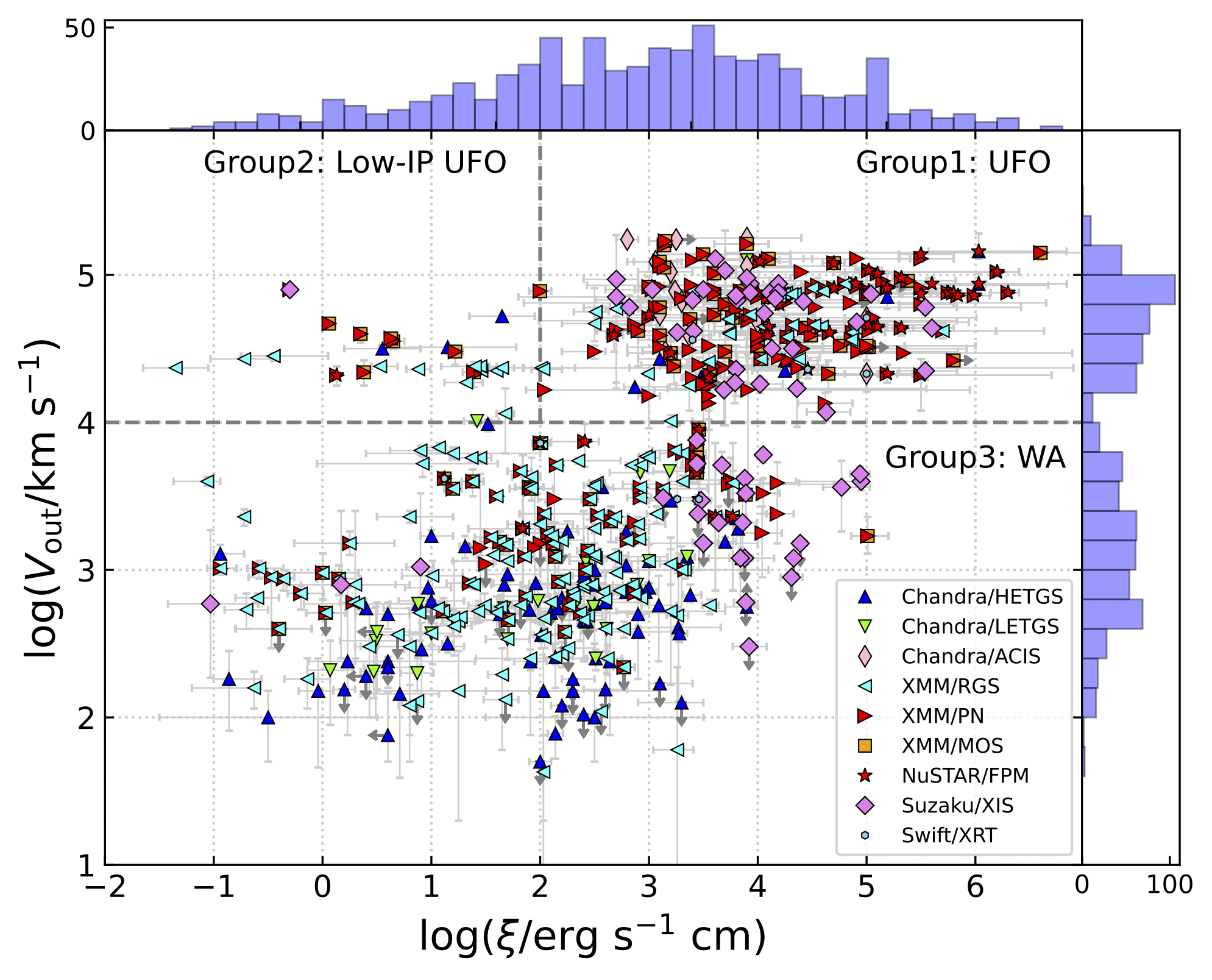}
    \includegraphics[keepaspectratio, scale=0.12]{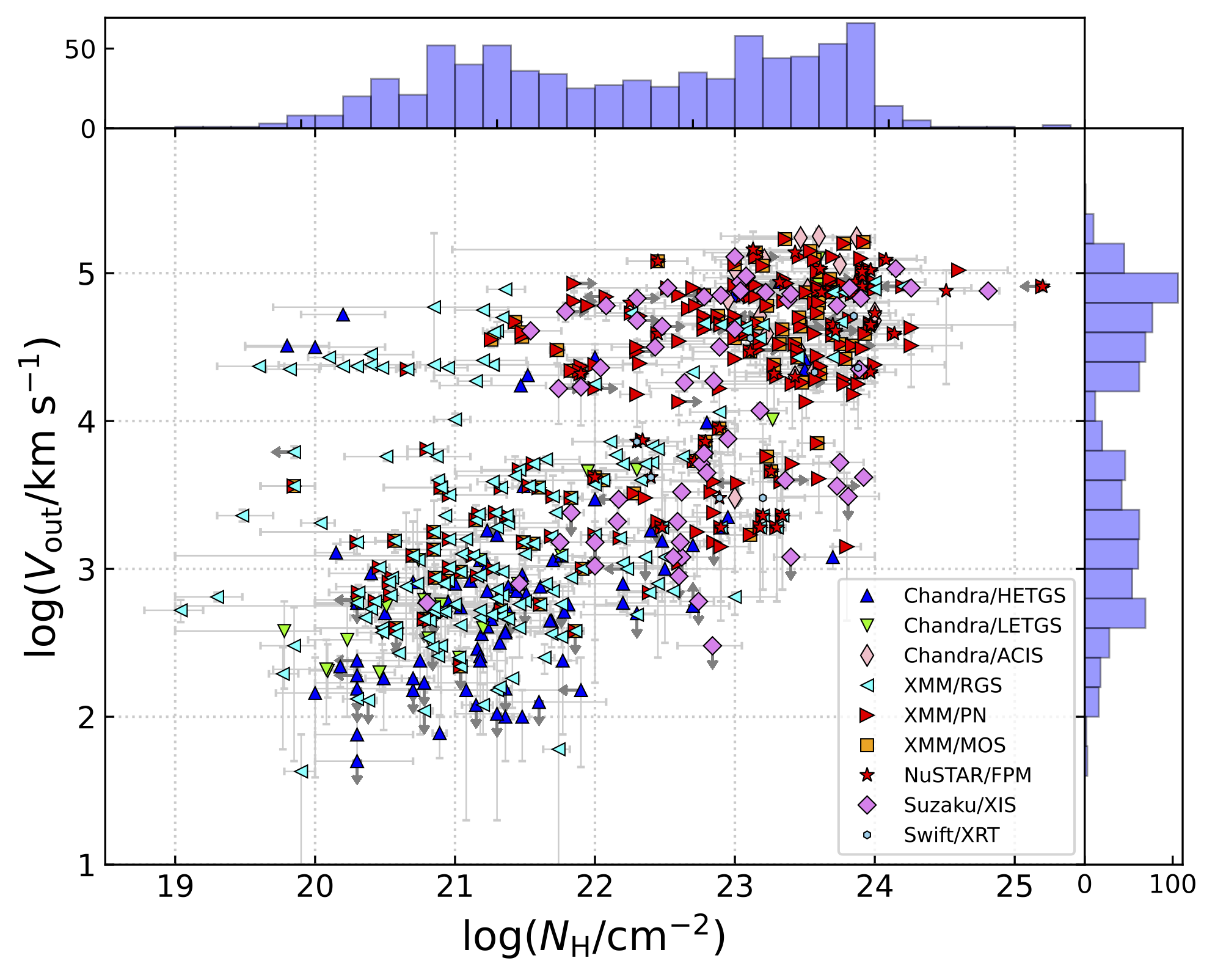}
    \includegraphics[keepaspectratio, scale=0.12]{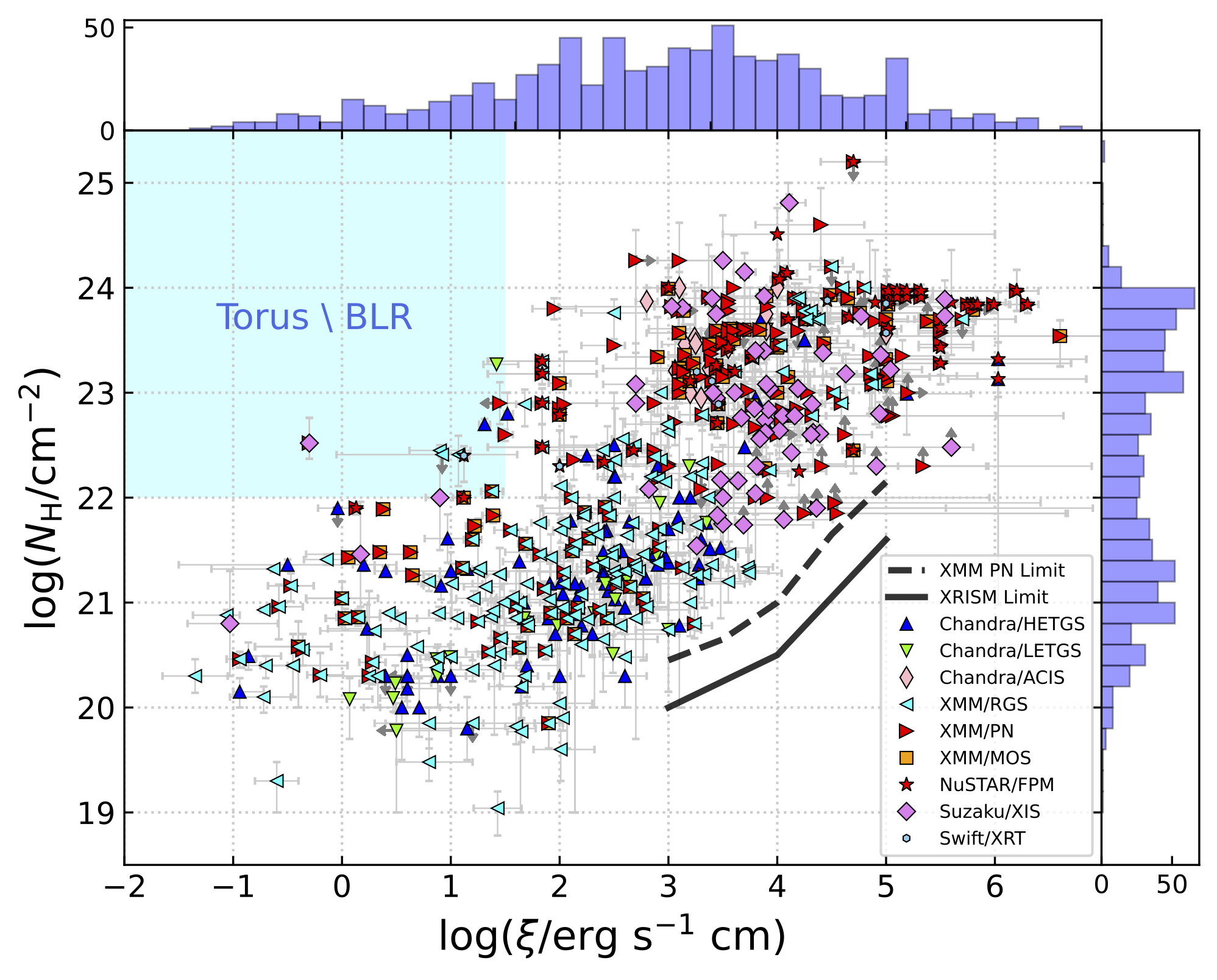}
    \caption{Same as in Figures~\ref{F3-outflow-opt-1} and top panels of Figure~\ref{F4-outflow-opt-2}, but symbols mark the instrument of observations of X-ray outflows.
    The dashed and solid curves indicate the simulated limitations of the 3$\sigma$ detection of the X-ray winds with XMM-Newton/PN and XRISM/Resolve, respectively.
    \\}
\label{F5-outflow-inst}
\end{figure*}

\subsection{Detection Biases} \label{sub3-2-bias}
To assess the observational biases of X-ray winds in the X-WING catalog attributed to instrumental performance (energy resolution and detection sensitivities), we categorized plots for the three parameters ($V_{\rm out}$, $\xi$, and $N_{\rm H}$) based on the X-ray instruments (refer to Figure~\ref{F5-outflow-inst}).
Following \cite{Kaastra2014WP}, the signal-to-noise (S/N) ratio for line detection is calculated as:
\begin{align} \label{Eq1-SN}
{\rm S/N} = \sqrt{A(E)tF(E)} \times W/\sqrt{{\rm max}(\Delta E, W)},
\end{align}
where $A(E)$ denotes the effective area, $t$ is the exposure time, $F(E)$ is the photon flux, $W$ represents the equivalent width (EW), and $\Delta E$ is the energy resolution of the instruments.
The value of $\sqrt{A(E)/\Delta E}$ ($\propto$ S/N for weak lines with $W \lesssim \Delta E$) mirrors the instruments' capability to detect absorption lines. Consequently, it is anticipated that weak lines are most easily detectable with Chandra/HETGS and XMM-Newton/RGS in the $\sim$1~keV band, and with XMM-Newton/PN, MOS, and Suzaku/XIS in the $\sim$6~keV band (refer to Appendix~\ref{AppendixA-inst} and Table~\ref{TA1-inst}).

For winds exhibiting ${\log}\xi \gtrsim 4$ or ${\log}N_{\rm H} \gtrsim 23$, the determination of small $V_{\rm out}$ values was restricted by XMM-Newton/PN, MOS, and Suzaku/XIS, with many of these values presented as upper limits (${\log}V_{\rm out} \lesssim 3$; refer to the top and bottom left panels of Figure~\ref{F5-outflow-inst}).
Using the equation for the line shift of an outflow in the rest-frame, $z_{\rm out} = \sqrt{(1+V_{\rm out}/c)/(1-V_{\rm out}/c)} - 1$ \citep[e.g.,][]{Danehkar2018c}, the line energies of \ion{Fe}{25} He$\alpha$ and \ion{Fe}{26} Ly$\alpha$ ($E_{\rm rest} = 6.70$--6.97~keV) are blueshifted to $E = 6.72$--6.99~keV if ${\log}(V_{\rm out}/{\rm km~s^{-1}}) \sim 3$. Given that some studies reported the presence of ionized absorbers with no velocity shift \citep[e.g.,][]{Blustin2005,Tombesi2013a,Laha2014b}, the observational biases could impact the detection of slower winds in the ${\log}V_{\rm out} \lesssim 3$ region (with ${\log}\xi \gtrsim 4$ or ${\log}N_{\rm H} \gtrsim 23$) due to the limited energy resolutions of XMM-Newton and Suzaku (130~eV; see Table~\ref{TA1-inst}).

\begin{figure*}
    \centering
    \includegraphics[keepaspectratio, scale=0.18]{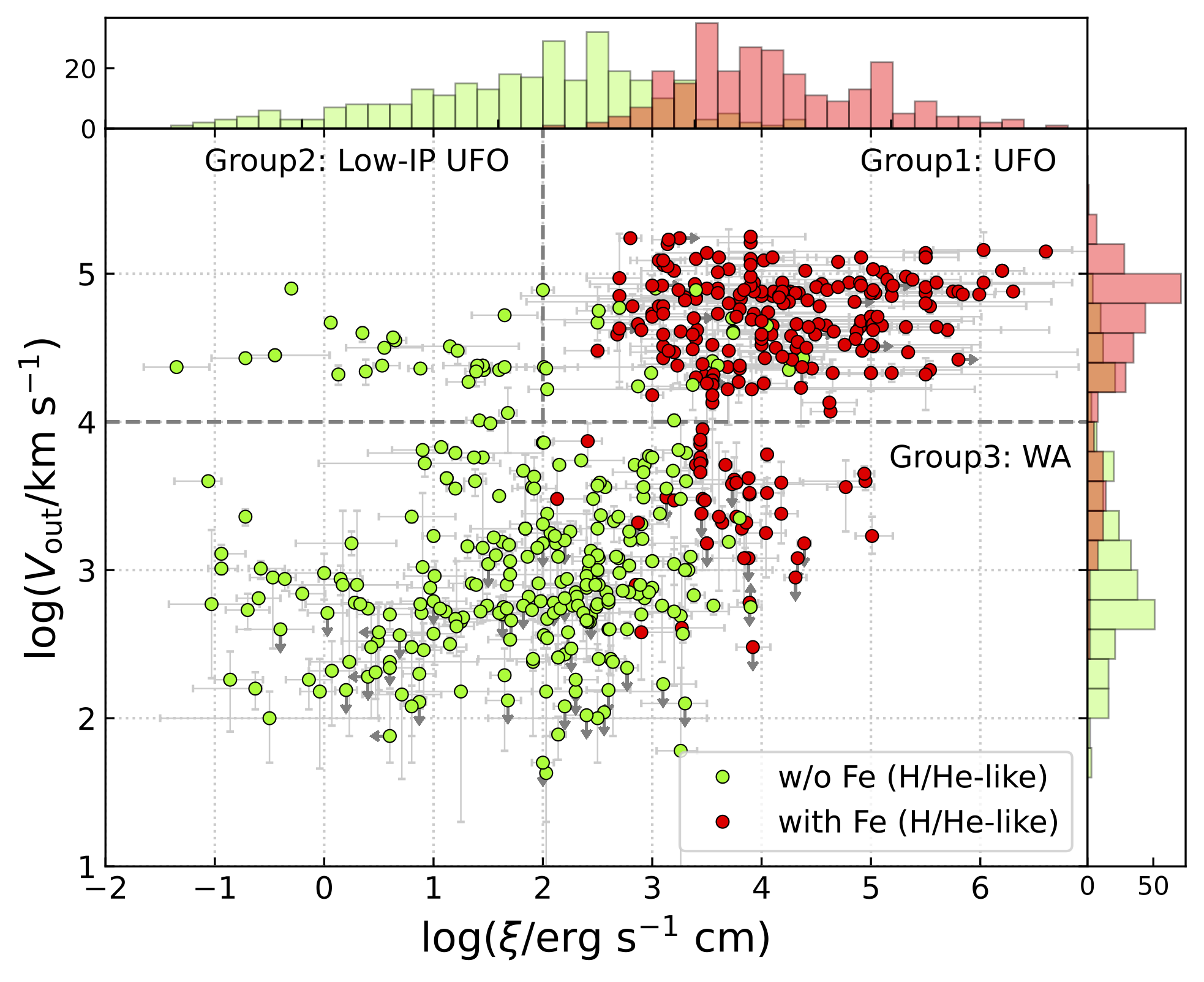}
    \includegraphics[keepaspectratio, scale=0.12]{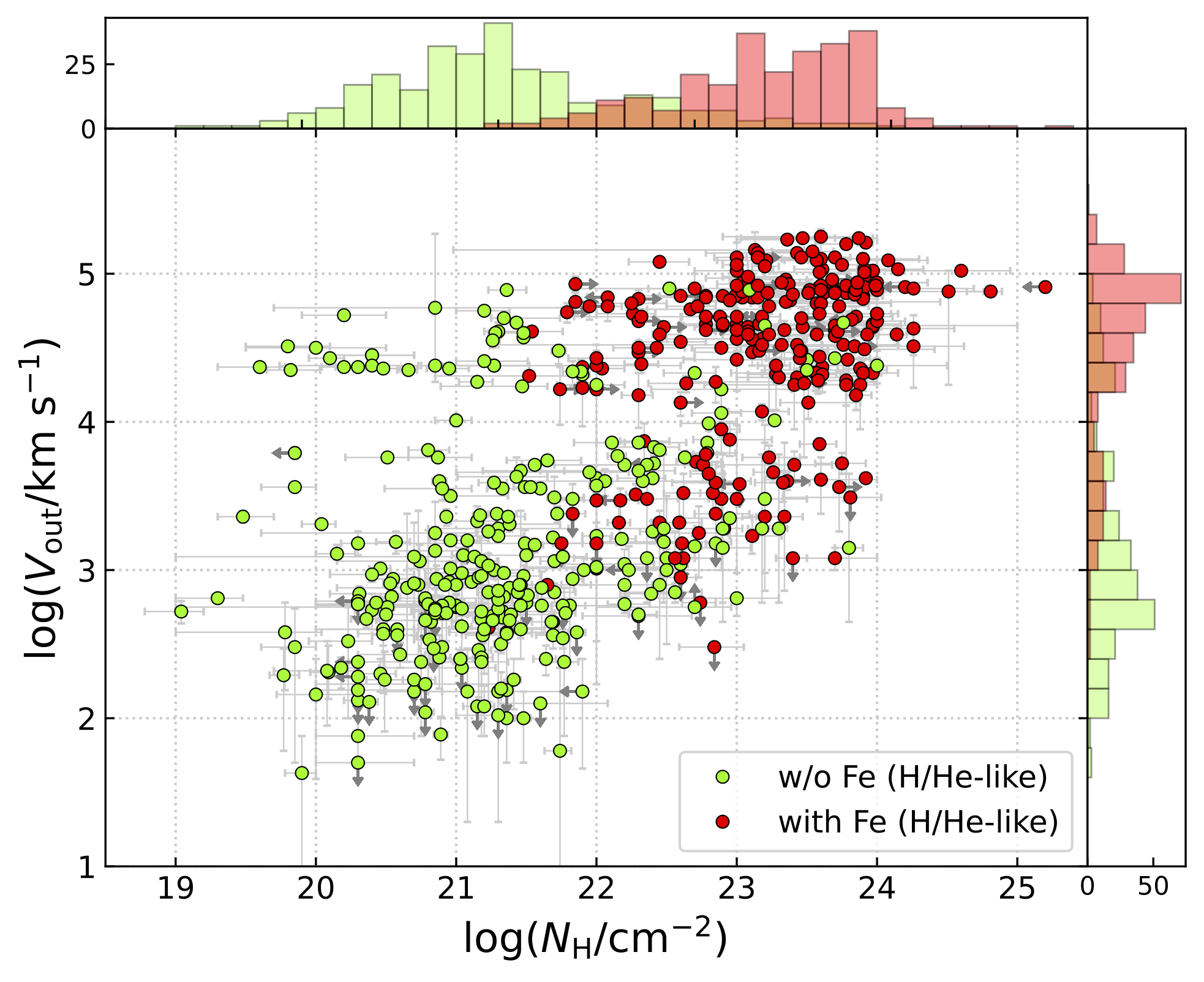}
    \includegraphics[keepaspectratio, scale=0.12]{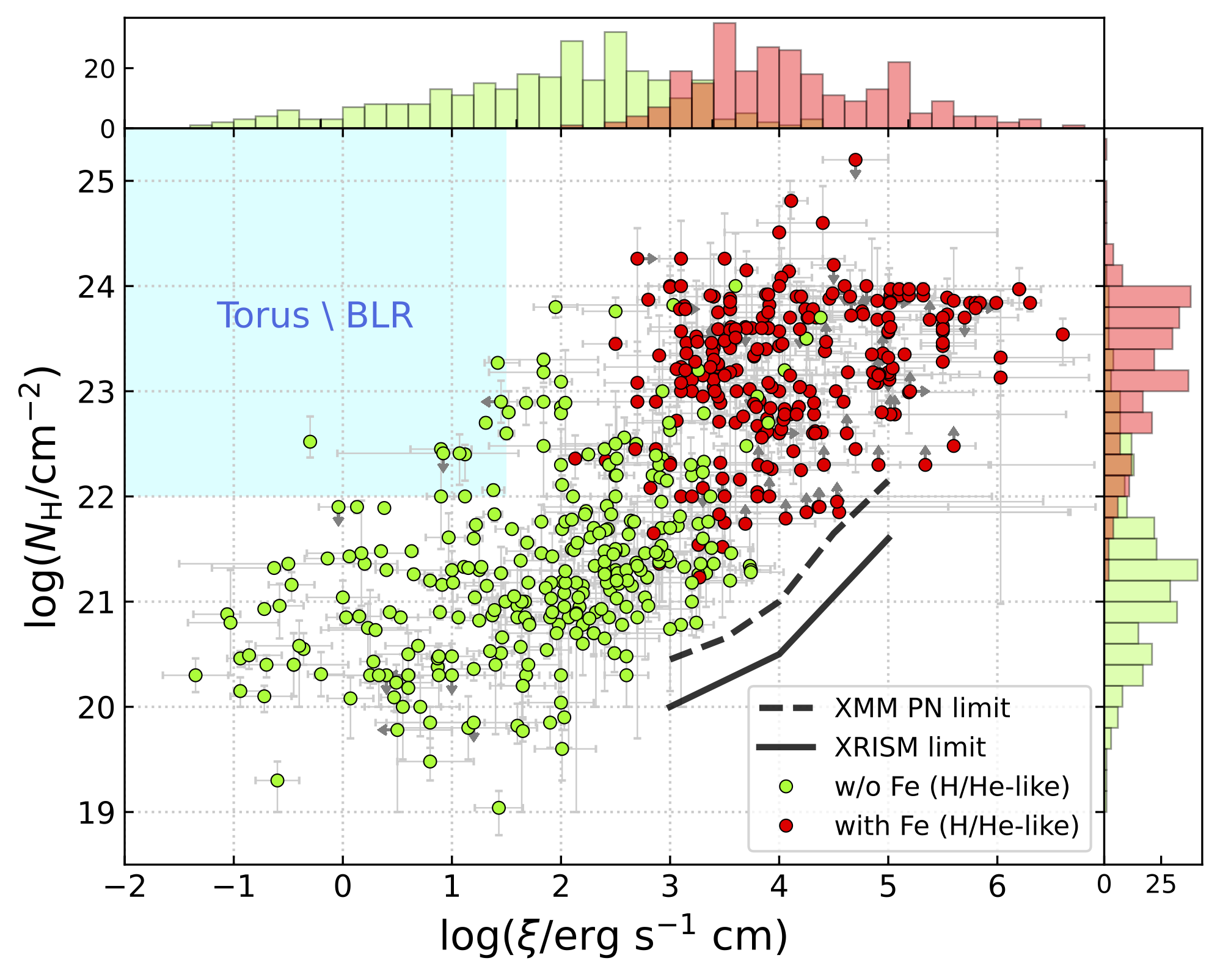}
    \caption{Same as in Figure~\ref{F5-outflow-inst}, but the symbols mark the methods of the detection of X-ray winds,
    utilized by the blueshifted Fe-K absorption lines \ion{Fe}{26} Ly$\alpha$ and \ion{Fe}{25} He$\beta$ above the rest-frame 6~keV band (red)
    or blueshifted absorption lines below the 6~keV band (light green).
    \\}
\label{F6-outflow-FeK}
\end{figure*}

In Section~\ref{sub3-1-winds}, we delve into potential biased regions in the ${\log}\xi$--${\log}N_{\rm H}$ plot. The winds located in the upper-left region (cyan-shaded area) in the bottom-right panel of Figure~\ref{F5-outflow-inst} pose challenges in detection due to soft X-ray obscuration caused by neutral absorbers. Conversely, the winds in the lower-right region are expected to be invisible, marked by weak lines with small $N_{\rm H}$.
To assess the detection limits, we computed the expected limits for ionized absorption lines using XMM-Newton/PN and compared them with simulations for the latest instrument, XRISM/Resolve. Given the complexity of spectral features in the soft X-ray band, arising from numerous unresolved absorption lines in the cyan region due to ionized absorbers with low $\xi$ and high $N_{\rm H}$, we specifically simulated detection limits for winds with high $\xi$ and low $N_{\rm H}$.
The S/N ratio of line detection was determined using Equation~\ref{Eq1-SN}. 
We adopted an instrumental performance ($A(E) = 210$~cm$^2$ and $\Delta E = 5$~eV) detailed in Appendix~\ref{AppendixA-inst} and assumed observational settings (exposure time and photon flux) based on a nearby ($D_{\rm L} = 19.0$~Mpc) Seyfert 1.5 galaxy NGC~4151, one of the brightest X-ray sources in the X-WING sample, selected as an early release target for XRISM observations. The 2--10~keV flux is $5 \times 10^{-11}~{\rm erg~s^{-1}~cm^{-2}}$ \citep{Ricci2017d}, and $t$ = 180~ksec. A typical photon index of 1.8 \citep{Ueda2014,Ricci2017d} was assumed for the conversion between the photon flux and the observed 2--10~keV flux.
The EWs of \ion{Fe}{25} He$\alpha$ and \ion{Fe}{26} Ly$\alpha$ in the ranges ${\log}\xi$ = 3--5 and ${\log}N_{\rm H}$ = 20--24 were simulated using the \textsc{xspec} model \textsf{mtable\{xout\_mtable.fits\}*powerlaw}. The \textsf{mtable\{xout\_mtable.fits\}} is the ionized absorption model generated by \citet{Ogawa2021} with \textsc{xstar} version 2.54a \citep{Kallman2001,Bautista2001}. The larger EWs from the two lines were utilized in each ${\log}\xi$ and ${\log}N_{\rm H}$ bin. Based on these values, we plotted the 3$\sigma$ detection limit curves using XMM-Newton/PN (dashed line) and XRISM/Resolve (solid line) in the bottom panel of Figure~\ref{F5-outflow-inst}. The limits of XMM-Newton/PN align with the plotted data (red triangles), suggesting that the X-ray winds reported in previous works are influenced by the detection bias. The detection limit curve (solid line) indicates a reduction in this bias with XRISM.

To summarize, our catalog reveals that observational biases have influenced the analysis of X-ray winds in two distinct ways. Firstly, slow winds (${\log}V_{\rm out} \lesssim 3$) with ${\log}\xi \gtrsim 4$ or ${\log}N_{\rm H} \gtrsim 23$ remain unexplored due to limitations in energy resolution. Secondly, the ${\log}\xi$--${\log}N_{\rm H}$ distribution solely represents the detectable region of X-ray winds \citep[also discussed by][]{Tombesi2013a}, aligning with expectations based on the instrumental performances of line detections (specifically, $\sqrt{A(E)/\Delta E}$).
These biases exhibit apparent correlations with the outflow parameters. Whether the existence of empty regions is attributed to detection bias or intrinsic characteristics, our data indicate that simple linear correlations cannot be universally applied to all targets.

The considerable dispersion observed among the outflow parameters ($V_{\rm out}$, $\xi$, and $N_{\rm H}$) may be attributed to their dependence on additional factors, such as inclination angles, AGN luminosities, and $M_{\rm BH}$. Some observations suggest clear correlations among the outflow parameters of specific objects \citep[as seen in][]{Pounds2013}. Given the diverse measurement methodologies employed in our catalog, further investigations into the systematic spectral analysis of X-WING sources using a standardized approach are essential to explore potential object-specific physical correlations.

\begin{figure*}
    \centering
    \includegraphics[keepaspectratio, scale=0.18]{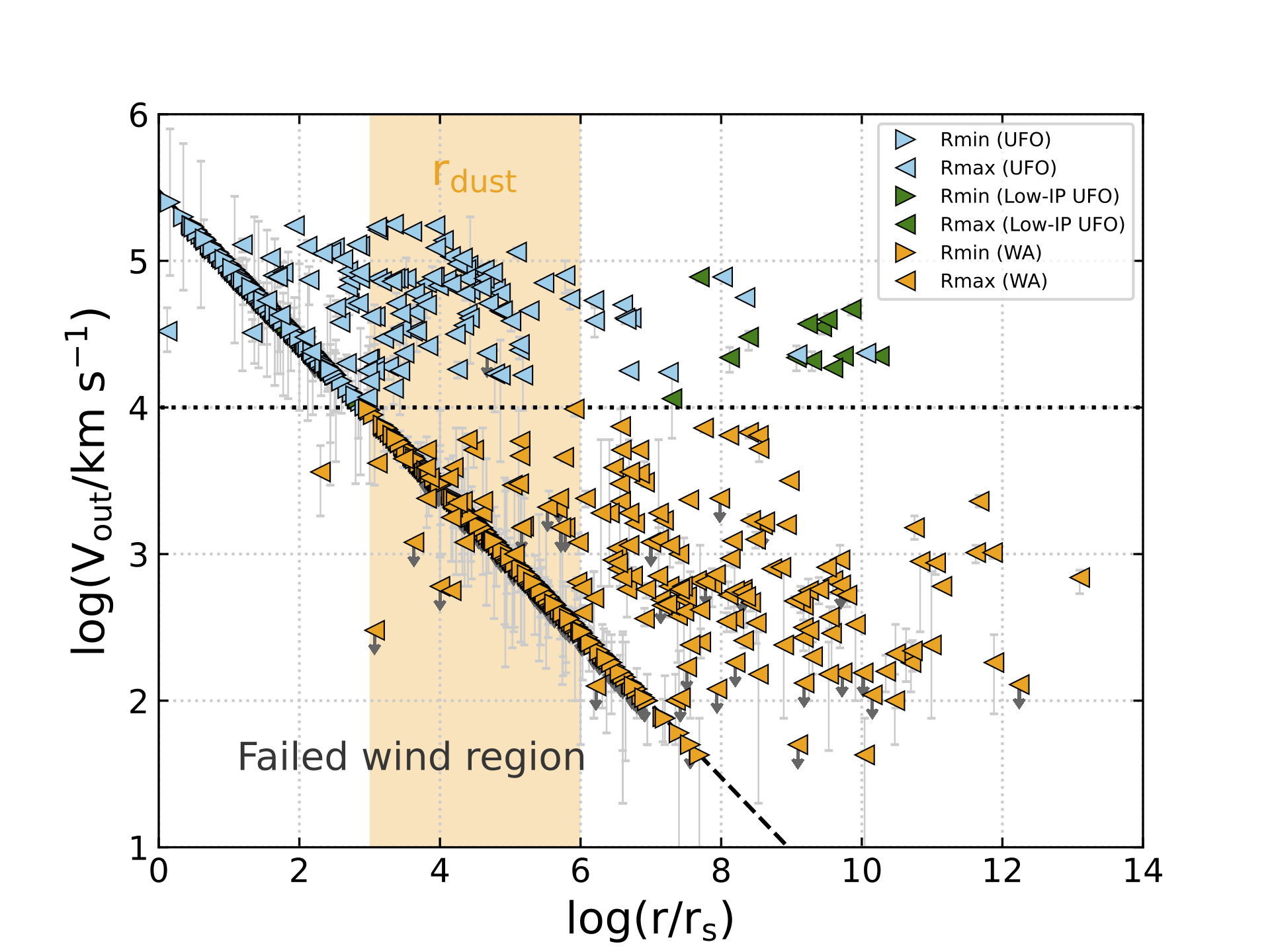}
    \includegraphics[keepaspectratio, scale=0.12]{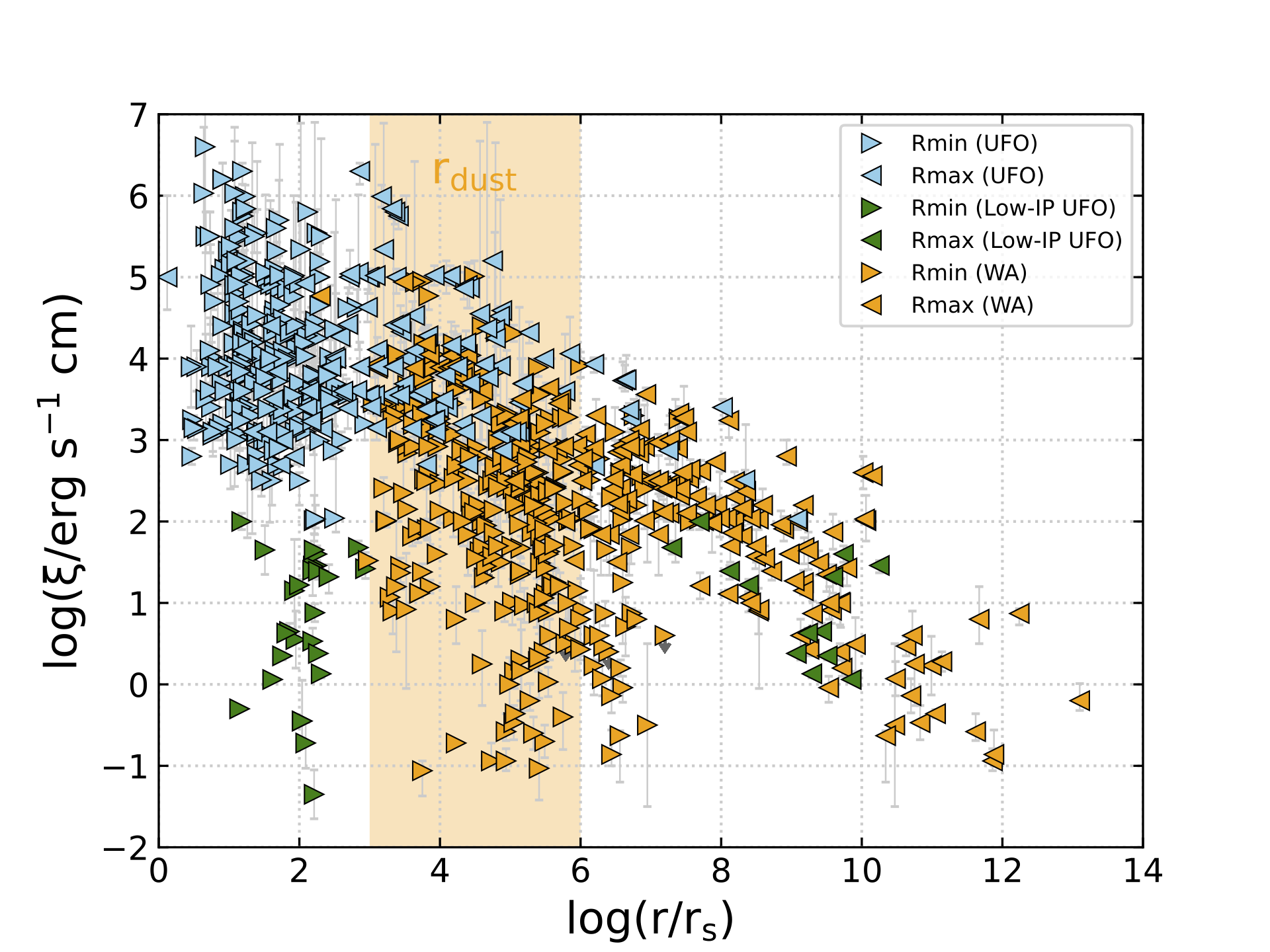}
    \includegraphics[keepaspectratio, scale=0.12]{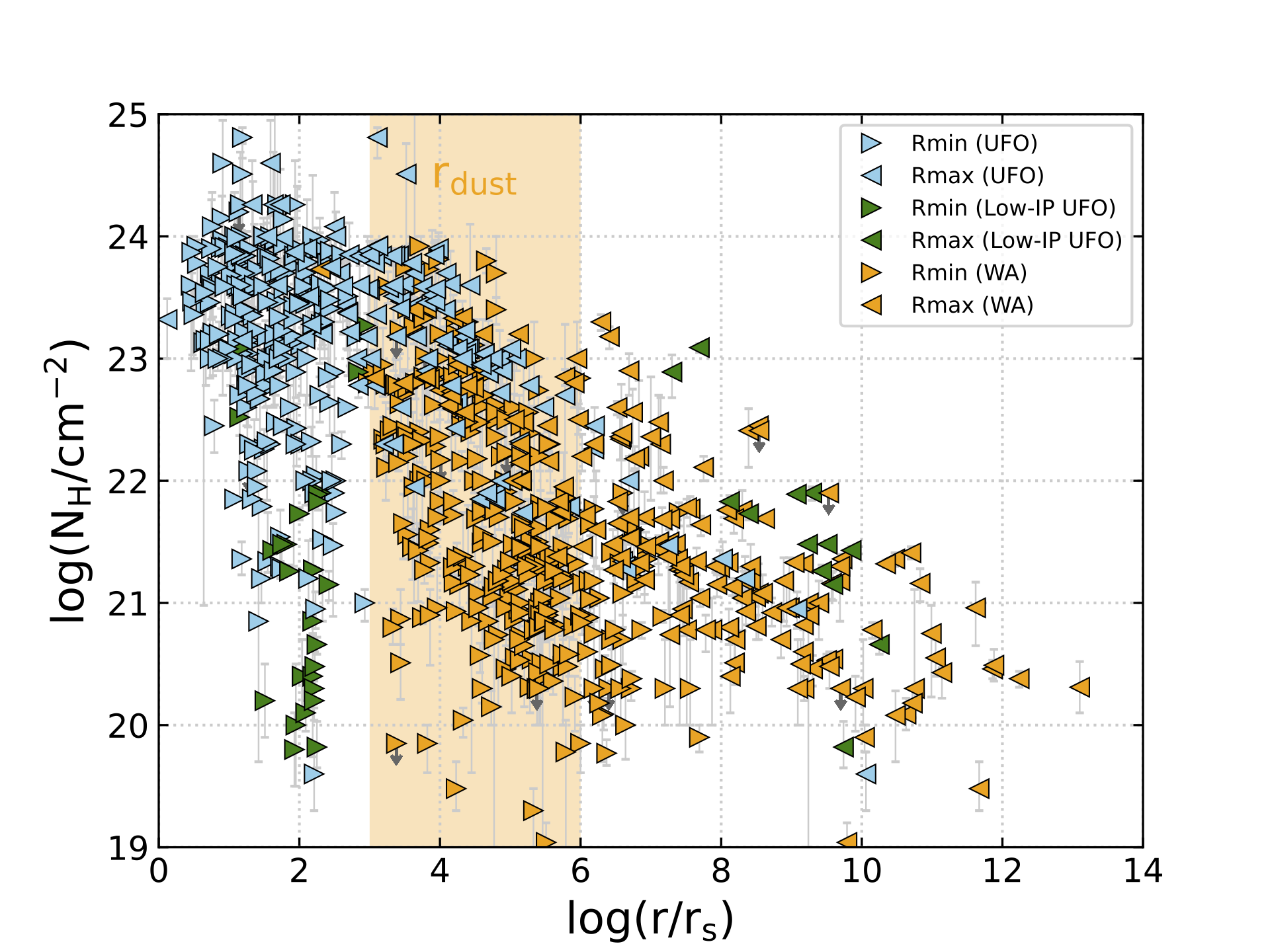}
    \caption{(Top) logarithmic $V_{\rm out}$ vs. logarithmic outflow radius normalized by Schwarzschild radius ($r/r_{\rm s}$).
    The dashed line shows the minimum outflow radius assuming the escape velocity, whereas the dotted line represents the threshold of the UFO and warm absorber.
    (bottom left) Logarithmic $\xi$ vs. log($r/r_{\rm s}$).
    (bottom right) Logarithmic $N_{\rm H}$ vs. log($r/r_{\rm s}$).
    The symbols mark the minimum or maximum outflow radii ($r = r_{\rm min}$ or $r_{\rm max}$) of the UFO (light blue), low-$\xi$ UFO (green), and warm absorber (orange).
    The orange-filled region indicates the sublimation radii of the X-WING AGNs.
    \\}
\label{F7-outflow-radius}
\end{figure*}

\subsection{Is the Velocity Gap Real?} \label{sub3-3-Fe}
We thoroughly examine the observational biases associated with the outflow parameters, taking into account both AGN types (Section~\ref{sub3-1-winds}) and instrumental constraints, namely, energy resolution and effective area (Section~\ref{sub3-2-bias}). Despite these considerations, our data, as depicted in Figures~\ref{F3-outflow-opt-1} and~\ref{F5-outflow-inst}, reveal a compelling phenomenon---a conspicuous gap occurring around $V_{\rm out} \sim 10,000~{\rm km~s^{-1}}$. This phenomenon remains unexplained by the aforementioned biases.
In Figure~\ref{F6-outflow-FeK}, we categorize X-ray winds based on the detection methods employed, specifically those utilizing \ion{Fe}{25}/\ion{Fe}{26} lines (above 6 keV). Strikingly, both methods, utilizing the \ion{Fe}{25}/\ion{Fe}{26} lines (depicted as red circles) and employing other lines in the $\lesssim$2~keV band (represented by light green circles), consistently indicate the existence of a gap around ${\sim}10,000~{\rm km~s^{-1}}$ (top and bottom left panels of Figure~\ref{F6-outflow-FeK}). The origin of this velocity gap between these two detection methods will now be explored.

For X-ray winds detected in the $\lesssim$2~keV band (or those without the \ion{Fe}{25}/\ion{Fe}{26} lines; represented by light green circles), the observed gap appears to be genuine. Numerous prior studies identified blueshifted lines, as exemplified by the low-$\xi$ UFOs detected with gratings such as Chandra/HETGS, LETGS, and XMM-Newton/RGS. These detections involved identifying multiple blueshifted lines in the rest-frame 0.5--2~keV band, originating from ions such as \ion{C}{5}, \ion{C}{6}, \ion{N}{4}, \ion{N}{5}, \ion{N}{6}, \ion{N}{7}, \ion{O}{5}, \ion{O}{6}, \ion{O}{7}, \ion{O}{8}, \ion{Ne}{7}, \ion{Ne}{8}, \ion{Ne}{9}, \ion{Ne}{10}, \ion{Si}{14}, \ion{Fe}{17}, \ion{Fe}{20}, and \ion{Fe}{21} \citep[e.g.,][]{Gupta2013c,Gupta2015,Pounds2014,Reeves2018a,Reeves2020,Sanfrutos2018,Krongold2021,XuYerong2021,XuYerong2022}. Despite potential complexities in the $A(E)$ functions of these grating arrays \citep[e.g.,][]{denHerder2001,Canizares2005}, the measurement of $V_{\rm out}$ through the identification of sets of blueshifted lines is likely less susceptible to such effects.
Given the absence of other artifacts in the $\lesssim$2~keV band, the observed velocity gap between low-IP UFOs and warm absorbers appears to be authentic. This suggests that the underlying physical mechanisms governing low-IP UFOs and warm absorbers are distinct. For instance, low-IP UFOs might be entrained as clumpy outflows at a scale of $\sim 100~{\rm pc}$ \citep[as proposed by][]{Serafinelli2019}, with velocities possibly reaching the theoretical limit determined by the terminal velocity from the accretion disk \citep{Luminari2021}. However, this scenario does not seem to be a plausible origin for warm absorbers (see Section~\ref{sub3-4-location}).

However, the methodology relying on \ion{Fe}{25}/\ion{Fe}{26} lines might be susceptible to artifacts. The rest-frame energies of the \ion{Fe}{25} He$\alpha$ (6.966~keV) and \ion{Fe}{26} Ly$\alpha$ (6.697~keV) lines with log$V_{\rm out}$ = 4 are 6.92~keV and 7.20~keV, respectively. Remarkably, the average of these two lines (7.06~keV) aligns precisely with the energy of the Fe K$\beta$ emission line (7.06~keV) originating from BLRs or tori. This alignment poses challenges for detecting distinct absorption lines with CCD detectors featuring a resolution of $\gtrsim$130~eV (refer to Table~\ref{TA1-inst}). Additionally, the presence of the Fe K-edge at 7.11~keV diminishes the significance of the blueshifted \ion{Fe}{26} Ly$\alpha$ absorption line. Such degeneracies make the identification of higher-$\xi$ winds with log$V_{\rm out}$ = 4 challenging, except in instances where deep absorption lines are detectable \citep{Reeves2010c,Giustini2011,Walton2018,Walton2020b}.
Given that the contribution of Fe~K$\beta$ line and Fe K-edge are less for unobscured AGNs, which are predominant in our sample (e.g., Seyfert 1.0--1.5), these potential artifacts might not be the primary source of these issues.
It is cautions that the detection of X-ray winds may be affected by the utilized instruments, models, and potential look-elsewhere effect (see Section~\ref{sub2-2-winds}).
If these artifacts and potential biases are not the main contributors, the spatial disconnection between UFOs and galactic-scale warm absorbers could be substantial, indicating a possible distinction in their driving mechanisms (refer to Section~\ref{sub3-4-location}).
This spatial disconnection holds significant implications for determining the reality of the $V_{\rm out}$ gap (i.e., disconnection), representing a crucial aspect in comprehending the AGN feedback from UFOs. 
Consequently, systematic surveys of X-ray winds utilizing XRISM are imperative to provide definitive answers.

\subsection{Locations of the X-ray Winds} \label{sub3-4-location}
Determining the location of X-ray winds holds significance in unraveling the mechanisms propelling them forward. 
Specifically, comparing the distances between the winds and the dust sublimation radius allows an assessment of their nature, distinguishing between dust-free and dusty gases \citep[e.g.,][]{Mehdipour2012}. 
Theoretically, owing to the considerably larger cross-section of dust compared to Thomson scattering, it is anticipated that radiation pressure from the AGN with $\lambda_{\rm Edd}$ surpassing the effective pressure ($\lambda_{\rm Edd}^{\rm eff} \sim 0.02$) can accelerate dusty gas outflows \citep{Fabian2008,Wada2015,Ishibashi2018}. 
Recent observational investigations indicate a significant decrease in the covering fractions of obscurers at $\lambda_{\rm Edd} \gtrsim \lambda_{\rm Edd}^{\rm eff}$ for Swift/BAT AGNs \citep[e.g.,][]{Ricci2017c,Ricci2022,Ricci2023d,Yamada2020,Ananna2022}, as well as heavily obscured AGNs in late mergers \citep[e.g.,][]{Yamada2019,Yamada2021,Yamada2024}. This lends support to the notion that dusty outflows possess sufficient strength to disperse circumnuclear material.
Moreover, considering the well-established impact of dust on the velocity dispersion gap between broad-line regions (BLRs) and narrow-line regions \citep{Netzer1993}, it might offer an explanation for the observed ${\log}V_{\rm out} \sim 4$ gap (Section~\ref{sub3-3-Fe}). 
While various plausible scenarios exist for the driving mechanisms of X-ray winds \citep[see e.g.,][]{Fukumura2022,Gallo2023}, we initially explore the possibility of radiation pressure on dusty gas being the primary driver for UFOs or warm absorbers.

The minimum and maximum distances of the X-ray winds ($r_{\rm min}$ and $r_{\rm max}$) can be estimated by using the equations in Section~2.1 of \citet{Gofford2015}:
\begin{align} \label{Eq2-3}
r_{\rm min} & = 2GM_{\rm BH}/V_{\rm out}^2 {\rm , and}\\
r_{\rm max} & = L_{\rm ion}/\xi N_{\rm H}.
\end{align}
The dust sublimation radius ($r_{\rm dust}$) at 1500~K can be calculated as
\begin{align} \label{Eq4}
r_{\rm dust} = 0.4 \times (L_{\rm bol,AGN}/10^{45}~{\rm erg~s^{-1}})^{1/2}~{\rm pc},
\end{align}
where $L_{\rm bol,AGN}$ denotes bolometric AGN luminosity \citep[e.g.,][]{Nenkova2008a,Nenkova2008b}.
We note that a dust-free outflow may exist at $r > r_{\rm dust}$ if it does not mix with the surrounding dust, whereas dust outflows are known to exist even in the pc region, as observed by IR interferometric observations \citep[e.g.,][]{Gamez-Rosas2022} and are expected to contain dust at $r > r_{\rm dust}$.
Here, we roughly estimated $r_{\rm dust}$ in our sample assuming $L_{\rm bol,AGN} = 20L_{\rm 2-10}$ \citep[e.g.,][]{Vasudevan2007} and obtained a range of ${\log}(r_{\rm dust}/r_{\rm s}) = 3.08$--6.06.
If adopting luminosity-dependent bolometric corrections derived from Equation~3 in \citet{Duras2020}, the values are almost unchanged, ${\log}(r_{\rm dust}/r_{\rm s}) = 3.03$--6.02.
This typical range ($\sim$3--6) aligns with the values derived from $L_{\rm bol,AGN}$ through UV-to-radio SED fitting, as detailed in Paper II. While Seyfert 2 galaxies are not prevalent in our sample (Section~\ref{sub3-1-winds}), it is essential to exercise caution when considering anisotropic radiation from the disk. In scenarios where such radiation is assumed, the dust sublimation radius ($r_{\rm dust}$) at the edge-on angle should theoretically be smaller than the values employed in our analysis, assuming isotropic radiation \citep[e.g.,][]{Czerny2011,Baskin2018,Kudoh2023}.

We conducted a comparative analysis of the distances of the X-ray winds (UFOs, low-IP UFOs, and warm absorbers) with the dust sublimation radii, as illustrated in Figure~\ref{F7-outflow-radius}.
In all the panels, the considerable disparities between $r_{\rm min}$ and $r_{\rm max}$ pose challenges in elucidating the relationship between the location and outflow parameters. In pursuing our primary objective, we specifically concentrate on comparing these radii with $r_{\rm dust}$ in this investigation. While the locations of low-IP UFOs exhibit minimal constraints, their scale is presumed to be extensive, as inferred from the variability of $V_{\rm out}$ and $N_{\rm H}$ (approximately 100~pc; \citealt{Serafinelli2019}), as previously discussed in Paper II.
The presence of outflows exhibiting $r_{\rm max} < r_{\rm min}$ indicates that either they are failed winds, or the wind stream line flows nearly perpendicular to the line of sight, resulting in the observation being limited to the tangential components of $V_{\rm out}$ \citep[e.g.,][]{Gofford2015}.
Typically, the locations of the UFOs exhibit $r_{\rm max} \lesssim r_{\rm dust}$, signifying that UFOs predominantly consist of dust-free gases.
The warm absorbers were situated at $r_{\rm min} \gtrsim r_{\rm dust}$.
However, considering that failed winds are present in dust tori \citep[e.g.,][]{Wada2012, Wada2016, Izumi2018} or dusty gas around the outer accretion disk \citep[e.g.,][]{Czerny2011, Baskin2018}, the locations of warm absorbers can be $r < r_{\rm dust}$ \citep[e.g.,][]{Reeves2017}. 
Thus, determining whether individual warm absorbers are dust-free (at $r < r_{\rm dust}$) or dusty (at $r > r_{\rm dust}$) becomes challenging. 
Independently, \citet{Mehdipour2018b} conducted high-resolution X-ray and mid-IR spectroscopy of IC~4329A, suggesting the presence of dust in the wind through the spectral shape of X-ray absorption (O, Si, and Fe edge features) \citep[e.g.,][]{Lee2001, Lee2013, Sako2003b}. 
Although XRISM may identify the absorption features of dust \citep[e.g.,][]{Ricci2023a, Vander-Meulen2023}, this technique is limited to the brightest AGNs. 
Alternatively, the variability timescales of the X-ray winds, systematically measured by reanalyzing approximately 20 years of observations (Table~\ref{TB4-obsid}) with a uniform model, could provide better constraints on their locations \citep[e.g.,][]{Serafinelli2019}. 
Furthermore, XRISM can efficiently detect the variability of UFOs due to its high sensitivity to line detection.

Our results suggest that UFOs are primarily composed of dust-free gas, challenging the notion that radiation pressure on dusty gas is the main driving mechanism.
There are two plausible models for the driving mechanisms of UFOs. The first is a radiation pressure-driven wind, where sufficiently high UV and soft X-ray
fluxes can accelerate weakly ionized plasma through enhanced force multipliers \citep[e.g.,][]{Proga2000, Sim2008, Hagino2015, Nomura2017, Mizumoto2021}. 
The second model involves an MHD-driven wind, where plasma is loaded onto a global poloidal magnetic field and accelerated by the Lorentz force \citep[e.g.,][]{Blandford1982, Ferreira1997, Fukumura2010, Fukumura2017, Jacquemin-Ide2021}.
XRISM can distinguish between these two models by observing the differences in their broad asymmetric UFO line spectra. For example, the former exhibits an extended red tail, and the latter shows a blue tail \citep[e.g.,][]{Fukumura2022}.
In contrast, warm absorbers would mainly be dusty winds if they did not fail at small scales. 
The radiation pressure on dusty gas can play a crucial role in launching winds from the dusty region, such as the torus or outer accretion disk. 
Notably, thermally driven winds due to Compton heating \citep[e.g.,][]{King2012,Mizumoto2017} can also launch winds from dusty regions. 
These warm absorber scenarios result in similar spectral features that are challenging to distinguish using X-ray observations alone.
A third possibility is that warm absorbers are shocked winds from the UFOs \citep[e.g.,][]{King2015}.
This scenario can be verified by testing whether the observed gap at $V_{\rm out} \sim 10,000~{\rm km~s^{-1}}$ is real or not. 
In addition, if the UFOs and warm absorbers are spatially disconnected, another question also arises as to where UFOs are blowing towards, such as BLRs or outer regions.
In conclusion, further studies employing XRISM are required to determine the most plausible mechanisms of UFOs and warm absorbers and comprehend the whole structure of X-ray winds.

\subsection{The Era of Calorimeter Spectrometers with XRISM} \label{sub3-5-xrism}
In conclusion, we reflect on the accomplishments of X-ray observations to date and anticipate new frontiers that XRISM will unveil in the near future. Through extensive X-ray observations and a plethora of studies, we have unveiled a greater diversity of X-ray winds than previously recognized. Notably, a crucial revelation is that $V_{\rm out}$, $\xi$, and $N_{\rm H}$ exhibit no clear correlation, albeit acknowledging that these results are currently influenced by instrumental limitations (discussed in Sections~\ref{sub3-2-bias} and~\ref{sub3-3-Fe}).

Owing to its high-energy resolution in the $\sim$6~keV band ($\Delta E \lesssim 5~{\rm eV}$), XRISM/Resolve reduces these observational biases.
Based on the simulations in this study, the XRISM solves the following problems:
\begin{enumerate}

    \item[(1)] XRISM/Resolve will detect \ion{Fe}{25}/\ion{Fe}{26} lines with small velocities thanks to the highest energy resolution ($\lesssim$5 eV) at 6~keV.
    Considering that its $E/\Delta E$ ($\sim 1000$) is comparable to that of Chandra/HETGS, 
    XRISM would reduce the biases for the slow winds from ${\log}V_{\rm out} \lesssim 3$ to $\sim$2 for high-$\xi$ and high-$N_{\rm H}$ values (Section~\ref{sub3-2-bias}).

    \item[(2)] The \ion{Fe}{25}/\ion{Fe}{26} lines with small EWs will be more detected with XRISM, particularly for the winds in the low-$\xi$/high-$N_{\rm H}$ and high-$\xi$/low-$N_{\rm H}$ regions, as described by the detection limit curve (Section~\ref{sub3-2-bias}).
    
    \item[(3)] Since XRISM observations are less affected by the artifacts, XRISM will solve whether the gap is real or not, 
    and furthermore, whether the UFOs can drive AGN feedback to the galaxies (Section~\ref{sub3-3-Fe}).

    \item[(4)] Its high sensitivity of line detections will help us to monitor the variability of UFOs and constrain their distances from the SMBH (Section~\ref{sub3-4-location}).

\end{enumerate}
For the simulation on the detection limit curve, we assumed $A(E) \geqslant 210$~cm$^2$ of the designed plan. 
The current status demonstrates ${\sim}180$~cm$^2$ even if the gate valve is closed (see Appendix~\ref{AppendixA1-inst}).
The operation of the gate valve will continue in the future, and if it opens, these anticipated results will be better.

Furthermore, XRISM is poised to unravel the intricacies of nuclear structures, encompassing accretion disks, BLRs, tori, UFOs, and warm absorbers by identifying their various emission/absorption lines and conducting narrow Fe K$\alpha$ reverberation mapping \citep[e.g.,][]{Zoghbi2019,Noda2023}.
It promises to offer novel insights into the driving mechanisms of X-ray winds (e.g., \citealt{Gallo2023} and the cited references). The work by \citet{Parker2022} emphasizes the existing degeneracy between reflection and wind components, heightening uncertainties in outflow properties. XRISM is anticipated to alleviate this degeneracy \citep[see also, e.g.,][]{Gallo2011d,Gallo2013c,Fabian2020,Middei2023}. With XRISM observations commencing in December 2023, updates are expected in the near future. Additionally, new challenges will surface, and the landscape will evolve with the advent of the New Advanced Telescope for High Energy Astrophysics (NewAthena), representing the next generation of X-ray telescopes.
\\

\section{Conclusion} \label{S4-summary}
This study marks the inaugural findings of the X-WING program, assembling a dataset comprising 132 AGNs with reported X-ray winds up to the conclusion of 2023. 
These winds were identified through the detection of blueshifted ionized absorption lines using instruments such as Chandra, XMM-Newton, NuSTAR, Suzaku, and Swift. 
Our X-WING database encompasses 573 X-ray winds, providing detailed parameters including $V_{\rm out}$, $\xi$, and $N_{\rm H}$. 
Additionally, we compile information on the OBSIDs, the applied X-ray fitting model, references, and relevant papers addressing duplicate reports.
The X-ray winds are categorized into three distinct outflow groups, UFOs, low-IP UFOs, and warm absorbers, based on the values of $V_{\rm out}$ and $\xi$. 
Despite the presence of low-IP UFOs, the comparison among $V_{\rm out}$, $\xi$, and $N_{\rm H}$ reveals the absence of straightforward linear correlations (Sections~\ref{S2-database} and~\ref{sub3-1-winds}).

The distributions of $V_{\rm out}$, $\xi$, and $N_{\rm H}$ can be reproduced using the detection limits anticipated using the instrumental performance ($A(E)$ and $\Delta E$).
This substantiates the notion that their distributions were subject to observational biases, even when employing the most extensive catalog available in the X-WING database.
We determined a velocity gap of approximately $V_{\rm out} \sim 10,000~{\rm km~s^{-1}}$.
The gap between the winds detected by the absorption lines in the $\lesssim$2~keV band suggested different origins for low-IP UFOs and warm absorbers.
Although the gap detected by the \ion{Fe}{25}/\ion{Fe}{26} lines can be due to confusion of the emission/absorption lines and the Fe K-edge, there is another possibility that the UFOs and galactic-scale warm absorbers might be disconnected (Sections~\ref{sub3-2-bias} and~\ref{sub3-3-Fe}).

The minimum and maximum radii of the winds suggest that UFOs are dust-free, whereas warm absorbers likely comprise dusty gas.
The location can be $r < r_{\rm min}$ for failed winds whose distances must be measured by their time variabilities.
From 2024, calorimeter spectrometers with XRISM will reduce biases and clarify the diversity of the winds. 
The XRISM determines whether the $V_{\rm out}$ gap between UFOs and warm absorbers is real, which is a key question for understanding AGN feedback from UFOs (Sections~\ref{sub3-4-location} and~\ref{sub3-5-xrism}).
\\

\textcolor{black}{We deeply appreciate the anonymous referee for very constructive comments and suggestions, which improved the quality of this manuscript.}
This study was financially supported by the
Japan Society for the Promotion of Science (JSPS) KAKENHI grant number 
22K20391, 23K13154 (S.Y.),
23K13153 (T.K.),
21K13958 (M.M.),
24K17104 (S.O.),
19K21884, 20H01941, 20H01947, 20KK0071 (H.N.),
20H01946 (Y.U.),
22K03683 (H.S.),
and 23KJ0450 (S.M.).
S.Y. and T.K. are grateful for support from the RIKEN Special Postdoctoral Researcher Program.
C.R. is grateful for the support from the Fondecyt Regular grant 1230345 and the ANID BASAL project FB210003.
This paper employs a list of Chandra datasets, obtained by the Chandra X-ray Observatory, contained in \dataset[DOI: 10.25574]{https://doi.org/10.25574/cdc.231}.

\vspace{5mm}
\facilities{NuSTAR, Chandra, XMM-Newton, Suzaku, Swift, XRISM.}
\software{HEAsoft (v6.31), \textsc{xspec} (v12.10.1; \citealt{Arnaud1996}), \textsc{sas} (v17.0.0; \citealt{Gabriel2004}).}

\appendix
\restartappendixnumbering

\section{X-ray Instruments and Models} \label{AppendixA-inst}
\subsection{X-ray Instruments}\label{AppendixA1-inst}
In this study, we focused on X-ray instruments that have been mainly utilized for identifying X-ray winds over the past 20 years.
The basic explanations of the X-ray instruments used in this study (see Section ~\ref{S2-database}) are as follows.

\begin{enumerate}
    \item Chandra/HETGS (\citealt{Canizares2005}).
    The HETGS consists of two sets of gratings: the Medium Energy Grating (MEG) and the High Energy Gratings (HEG).
    The grating provides spectral resolving powers of up to 1000 over the range 0.4–8 keV.

    \item Chandra/LETGS (\citealt{Brinkman2000}).
    The LETGS is of a similar design to the HETGS but is optimized for energies less than 1 keV.

    \item Chandra/ACIS (\citealt{Garmire2003}).
    ACIS is arranged as an array of ten CCDs, one as a two by two array (ACIS-I), and one as a one by six array (ACIS-S).
    The two CCDs on ACIS-S are back-illuminated (BI) and the others on ACIS-I and ACIS-S are front-illuminated (FI).

    \item XMM-Newton/RGS (\citealt{denHerder2001}), consisting of two identical RGS grating arrays (RGS1 and RGS2).

    \item XMM-Newton/EPIC PN and MOS.
    There are three EPIC detectors; one contains twelve PN CCDs \citep{Struder2001}, while the two use seven MOS CCDs (MOS1 and MOS2; \citealt{Turner2001}).

    \item NuSTAR/FPM (\citealt{Harrison2013}), consisting of two coaligned X-ray telescopes coupled with FPMA and FPMB.
    
    \item Suzaku/XIS (\citealt{Mitsuda2007}), consisting of four X-ray CCD cameras, XIS-0, XIS-1, XIS-2, and XIS-3.
    The XIS-0, 2, 3 are FI cameras and XIS-1 is BI.
    The XIS-2 data were not available after 2006 November because the malfunction had occurred.
    
    \item Swift/XRT (\citealt{Evans2009}), a CCD imaging spectrometer designed to follow-up the X-ray transients and variable sources.
    
\end{enumerate}
For Chandra, HETGS is most commonly used with ACIS-S, whereas LETGS is used with ACIS-S and a High Resolution Camera spectroscopic array (HRC-S) \citep[e.g.,][]{Brinkman2000,Canizares2005}.
In Table~\ref{TA1-inst}, we summarize the basic performance of the energy bands, wavelength ranges, energy resolutions ($\Delta E$), spectral resolving powers ($E$/$\Delta E$), effective areas $A$($E$), ratios of $A$($E$)/$\Delta E$, and references.

In September 2023, the X-ray observatory XRISM \citep{Tashiro2018} was launched.
The XRISM payload comprises two instruments: Resolve and Xtend.
Resolve is an X-ray microcalorimeter with a 6-by-6-pixel detector, which is capable of high-energy-resolution spectroscopy.
Xtend, an array of four CCD detectors, has the imaging-spectroscopic capability of a wide FoV ($38 \times 38$~arcmin$^{2}$) and medium energy resolution.
Based on the ground calibration, the expected performances were as follows: in the 6~keV band, $A(E) \geqslant 210~{\rm cm}^2$ and $\Delta E \leqslant 7$~eV for Resolve 
and $A(E) \geqslant 300~{\rm cm}^2$ and $\Delta E \leqslant 250$~eV for Xtend.\footnote{\url{https://xrism.isas.jaxa.jp/research/analysis/manuals/xrqr_v1.pdf}}
According to the recent XRISM first-light observations \footnote{\url{https://heasarc.gsfc.nasa.gov/docs/xrism/}}, Resolve reaches a resolution of 5~eV.
However, the gate valve of the Resolve instrument (X-ray aperture door) did not open in January 2024, which lowered the effective area.\footnote{\url{https://heasarc.gsfc.nasa.gov/docs/xrism/proposals/index.html}}
The public ancillary response files of Resolve for GO Cycle 1 proposers demonstrate $A(E) \sim 180~{\rm cm}^2$ assuming a closed gate valve.
To make it easier to apply the simulated results in both cases of the gate valve, we adopted a typical value of $A(E) = 210~{\rm cm}^2$ and $\Delta E = 5$~eV for the spectral simulation of Resolve (Section~\ref{sub3-3-Fe}).

\begin{deluxetable*}{llccccccc}
\label{TA1-inst}
\tablecaption{X-ray Instruments Utilized in the Previous Works for Detecting X-ray Winds}
\tabletypesize{\footnotesize}
\tablehead{
\colhead{Facility} &
\colhead{Instrument} &
\colhead{Band} &
\colhead{Wavelength} &
\colhead{$\Delta E$} &
\colhead{$E$/$\Delta E$} &
\colhead{$A$($E$)} &
\colhead{$A$($E$)/$\Delta E$} &
\colhead{Ref.} \\
 & & (keV) & (\AA) & (eV) & & (cm$^{2}$) & (cm$^{2}$/eV) &
}
\startdata
\multicolumn{9}{c}{Performance at the 1~keV Band} \\
\hline
Chandra    & HETGS (w/ ACIS-S)            & 0.4--10   & 1.2--31   &   1 & 1000 &   59  & 59.0 & 1,2    \\
           & LETGS (w/ ACIS-S)            & 0.2--10   & 1.2--62   &   4 &  250 &   23  &  5.8 & 1,3    \\
           & LETGS (w/ HRC-S)             & 0.07--10  & 1.2--175  &   4 &  250 &   24  &  6.0 & 1,3    \\
           & ACIS-I                       & 0.2--10   & 1.2--62   &  60 &   17 &  367  &  6.1 & 1,4    \\
           & ACIS-S (FI)                  & 0.2--10   & 1.2--62   &  60 &   17 &  367  &  6.1 & 1,4    \\
           & ACIS-S (BI)                  & 0.2--10   & 1.2--62   & 120 &    8 &  555  &  4.6 & 1,4    \\
XMM-Newton & RGS ($\times$2; RGS1+RGS2)   & 0.35--2.5 & 5.0--35   &   3 &  333 &  185  & 61.7 & 5,6,7  \\
           & PN                           & 0.15--12  & 1.0--83   &  70 &   14 & 1227  & 17.5 & 5,8    \\
           & MOS ($\times$2; MOS1+MOS2)   & 0.15--12  & 1.0--83   &  80 &   13 & 1844  & 23.1 & 5,8    \\
Suzaku     & XIS ($\times$4; XIS 0+1+2+3) & 0.2--12   & 1.0--62   &  50 &   20 & 1100  & 22.0 & 6,9,10 \\
\hline
\multicolumn{9}{c}{Performance at the 6~keV Band} \\
\hline
Chandra    & HETGS (w/ ACIS-S)            & 0.4--10   & 1.2--31   &  29 & 207 &   25 &  0.9 & 1,2,6 \\
           & LETGS (w/ ACIS-S)            & 0.2--10   & 1.2--62   & 150 &  40 &   20 &  0.1 & 1,3,6 \\
           & LETGS (w/ HRC-S)             & 0.07--10  & 1.2--175  & 150 &  40 &    4 &  0.03& 1,3,6 \\
           & ACIS-I                       & 0.2--10   & 1.2--62   & 130 &  46 &  235 &  1.8 & 1,4,6 \\
           & ACIS-S (FI)                  & 0.2--10   & 1.2--62   & 130 &  46 &  235 &  1.8 & 1,4,6 \\
           & ACIS-S (BI)                  & 0.2--10   & 1.2--62   & 170 &  35 &  205 &  1.2 & 1,4,6 \\
XMM-Newton & PN                           & 0.15--12  & 1.0--83   & 130 &  46 &  851 &  6.5 & 5,6   \\
           & MOS ($\times$2; MOS1+MOS2)   & 0.15--12  & 1.0--83   & 130 &  46 & 1536 & 11.8 & 5,6   \\
NuSTAR     & FPM ($\times$2; FPMA+FPMB)   & 3--79     & 0.16--4   & 400 &  15 &  800 &  2.0 & 11,12 \\
Suzaku     & XIS ($\times$4; XIS 0+1+2+3) & 0.2--12   & 1.0--62   & 130 &  46 & 1000 &  7.7 & 9,10  \\
Swift      & XRT                          & 0.2--10   & 1.2--62   & 140 &  43 & 20$^{\dagger}$ & $\sim$0.1 & 13 \\
\enddata
\tablecomments{References:
(1) Chandra Pocket Guide (\url{https://cxc.harvard.edu/cdo/pocket_guide.pdf}).
(2-3) Chapters 8 and 9 of The Chandra Proposer's Observatory Guide (\url{https://cxc.harvard.edu/proposer/POG/html/index.html}).
(4) HEASARC for Chandra (\url{https://heasarc.gsfc.nasa.gov/docs/chandra/chandra.html}).
(5) XMM-Newton Users Handbook (\url{https://xmm-tools.cosmos.esa.int/external/xmm_user_support/documentation/uhb/basics.html}).
(6) HEASARC (\url{https://heasarc.gsfc.nasa.gov/docs/heasarc/missions/comparison.html})
(7) XMM-Newton Calibration technical notes (\url{https://xmmweb.esac.esa.int/docs/documents/CAL-TN-0030.pdf})
(8) HEASARC for XMM-Newton (\url{https://heasarc.gsfc.nasa.gov/docs/xmm/xmm.html})
(9) The Suzaku Web Page (\url{https://space.mit.edu/XIS/about/}).
(10) \citet{Mitsuda2007};
(11) NuSTAR Observatory guide (\url{https://heasarc.gsfc.nasa.gov/docs/nustar/nustar_obsguide.pdf})
(12) \citet{Harrison2013};
(13) The SWIFT XRT data reduction guide (\url{https://swift.gsfc.nasa.gov/analysis/xrt_swguide_v1_2.pdf}).
For Swift/XRT, the symbol $\dagger$ indicates $A$($E$) at 8.1~keV.
}
\end{deluxetable*}

\subsection{X-ray Models of the Outflowing Ionized Absorbers}\label{AppendixA2-models}
Various models for outflowing ionized absorbers have been employed in the literature, with a comprehensive summary provided in Table~\ref{TB3-outflow}. 
The \textsc{xstar} code, introduced by \citet{Kallman1996,Kallman2001,Kallman2004}, played a prominent role in generating models for photoionized absorbers, accounting for 267 out of 573 analyzed winds. 
Studies have utilized \textsc{xstar} photoionization models to characterize outflowing ionized absorbers without scattering (\textsf{warmabs}), those with scattering off the outflowing absorber (\textsf{windabs}), and scenarios involving partial covering of partially ionized absorbing material (\textsf{zxipcf}; \citealt{Reeves2008}).
Additionally, \citet{Danehkar2018d} introduced the \textsc{mpi\_xstar} model, a parallel implementation involving multiple \textsc{xstar} runs. Another simulation code, \textsc{cloudy} \citep{Ferland1998,Ferland2013}, was also enlisted for photoionization simulations.
In \textsc{spex} \citep{Kaastra1996}, the \textsf{xabs} model, a photoionized absorption model, was prominently employed in numerous studies, encompassing 130 out of 573 examined winds. In this approach, the photoionization equilibrium was pre-calculated for a grid of $\xi$ values using an external code (such as \textsc{xstar} or \textsc{cloudy}). 
Conversely, the photoionized absorption/emission model \textsf{pion} within \textsc{spex} performs self-consistent calculations of photoionization equilibrium using available plasma routines.
Note that these models, specifically \textsc{xstar} and \textsc{cloudy}, have undergone multiple updates. Some investigations have opted for the PHotoionized Absorption Spectral Engine (\textsc{phase}; \citealt{Krongold2003}), employing a model assuming a geometry featuring a central source emitting an ionizing continuum, interacting with clouds of gas computed using \textsc{cloudy}. Additionally, \textsc{absori} \citep{Done1992,Zdziarski1995}, the \textsc{xspec} model for ionized absorbing plasma, has found utility in sources displaying absorption edges.

Additionally, the absorption measure distribution (AMD; \citealt{Holczer2007}) method was adopted to reconstruct the actual distribution of the column density in the plasma as a continuous function of $\xi$.
\citet{Igo2020} calculated the fractional excess variance (FEV; \citealt{Vaughan2003}) spectra to search for UFOs (Section~\ref{sub2-1-sample}).
Some studies have identified absorption line candidates using the Gaussian model or by judging from the residual in the spectral fittings.
\\

\section{Database of X-WING Sample and X-ray Winds} \label{AppendixB-database}
The main information on the X-WING Sample (Section~\ref{sub2-1-sample}) is summarized in Table~\ref{TB2-sample}.
The X-ray wind database (section ~\ref{sub2-2-winds}) is presented in Tables~\ref{TB3-outflow} and~\ref{TB4-obsid}.

\clearpage

\startlongtable
\begin{deluxetable*}{llcccccccccc}
\label{TB2-sample}
\tablecaption{Target Information in the X-WING Sample}
\tabletypesize{\scriptsize}
\tablehead{
\colhead{ID} &
\colhead{Object} &
\colhead{$z$} &
\colhead{$D_{\rm L}$} &
\colhead{Class} &
\colhead{$F_{\rm 2-10}$} &
\colhead{log$L_{\rm 2-10}$} &
\colhead{log$M_{\rm BH}$} &
\colhead{Method} &
\colhead{$N_{\rm w}$} &
\colhead{UFO} &
\colhead{References (3-4,5,6-7,8-9)} 
}
\decimalcolnumbers
\startdata
\multicolumn{12}{c}{Nearby (z$<$1) Galaxies} \\
\hline
1 & 1E~0754.6+3928 & 0.0960 & 440.7 & NLS1 & 2.2 & 43.70 & 8.02 & RM & 4 & Y & (Md20;Md20;Se07;Md20) \\
2 & 1ES~1927+654 & 0.0194 & 84.4 & 2.0 & 3.7 & 42.50 & 6.00 & Ms & 2(*) & n & (Ri20;VC10;Ri20;Ga13) \\
3 & 1H~0323+342 & 0.0630 & 282.7 & NLS1 & 0.9 & 42.93 & 6.76 & Hb & 2(*) & n & (Me19;Pl14;K22A;Ri17) \\
4 & 1H~0419-577 & 0.1040 & 480.0 & 1.5 & 14.3 & 44.60 & 8.34 & Hb & 1 & Y & (To10;VC10;K22A;Ri17) \\
5 & 1H~0707-495 & 0.0406 & 179.3 & NLS1 & 1.1 & 42.66 & 6.31 & Hb & 9 & Y & (Xu21;Xu21;Bi09;Bi09) \\
6 & 1H~1934-063 & 0.0102 & 44.7 & NLS1 & 23.2 & 42.75 & 6.61 & Hb & 2 & n & (Xu22;Xu22;K22A;Ri17) \\
7 & 2MASSI~J0918486+211717 & 0.1490 & 708.0 & 1.5 & 1.2 & 43.89 & 7.35 & Hb & 1 & Y? & (Po07;VC10;Sh11;Bi09) \\
8 & 2MASS~J1051+3539 & 0.1590 & 760.2 & 1.9 & 1.7 & 44.15 & 8.40 & Hb & 1 & Y & (Ma23;VC10;Ma23;Bi09) \\
9 & 2MASS~J1653+2349 & 0.1030 & 475.1 & 2.0 & 2.0 & 44.06 & 8.17 & Vdisp & 1 & Y & (Ma23;VC10;K22B;Ri17) \\
10 & 3C~105 & 0.0890 & 406.6 & 2.0 & 2.2 & 44.30 & 8.59 & Vdisp & 1 & Y & (To14;To14;K22B;Ri17) \\
11 & 3C~111 & 0.0485 & 215.4 & 1.0 & 55.5 & 44.53 & 8.45 & Hb & 3 & Y & (To14;To14;K22A;Ri17) \\
12 & 3C~120 & 0.0330 & 144.9 & 1.5 & 36.9 & 43.98 & 7.74 & RM & 4 & Y & (Me19;Me19;Be15;Ri17) \\
13 & 3C~273 & 0.1580 & 755.0 & 1.0 & 117.8 & 45.88 & 8.84 & RM & 1(*) & n & (Me19;Me19;Be15;Ri17) \\
14 & 3C~382 & 0.0580 & 259.3 & 1.0 & 37.8 & 44.60 & 8.85 & RM & 5 & n & (Me19;Me19;Be15;Ri17) \\
15 & 3C~390.3 & 0.0560 & 250.0 & 1.5 & 42.6 & 44.52 & 8.71 & RM & 4 & Y & (Me19;Me19;Be15;Ri17) \\
16 & 3C~445 & 0.0558 & 249.1 & 1.5 & 6.4 & 44.25 & 7.89 & Hb & 2 & n & (To14;To14;K22A;Ri17) \\
17 & 3C~59 & 0.1110 & 514.7 & 1.8 & 8.2 & 44.42 & 8.90 & Ha & 2(*) & n & (Me19;Me19;Si07;Ri17) \\
18 & 4C~+31.63 & 0.2980 & 1540.8 & 1.0 & 4.1 & 45.07 & 8.52 & Hb & 1(*) & n & (Me19;Me19;K22A;Ri17) \\
19 & 4C~+34.47 & 0.2060 & 1013.0 & 1.0 & 8.5 & 45.02 & 8.30 & Hb & 1 & n & (Me19;Me19;K22A;Ri17) \\
20 & 4C~+74.26 & 0.1040 & 480.0 & 1.0 & 24.4 & 44.87 & 9.83 & Hb & 5 & Y & (Me19;Me19;K22A;Ri17) \\
21 & Ark~120 & 0.0327 & 143.6 & 1.0 & 27.0 & 43.81 & 8.07 & RM & 1 & Y & (To10;VC10;Be15;Ri17) \\
22 & Ark~564 & 0.0240 & 104.7 & NLS1 & 16.3 & 43.34 & 6.27 & Hb & 11 & Y & (La14;Ni09;Bi09;Bi09) \\
23 & CBS~126 & 0.0789 & 358.0 & 1.2 & 2.2 & 43.53 & 7.93 & Hb & 1 & n & (NED;VC10;Ch19;Ch19) \\
24 & Centaurus~A & 0.0018 & 3.7 & 2.0 & 184.9 & 42.39 & 8.29 & Vdisp & 3 & n & (To14;To14;K22B;Ri17) \\
25 & ESO~075-41 & 0.0280 & 122.5 & 1.0 & 6.1 & 43.04 & 8.07 & Hb & 1(*) & n & (Me19;Me19;K22A;Ri17) \\
26 & ESO~103-35 & 0.0130 & 56.2 & 2.0 & 22.3 & 43.36 & 7.37 & Vdisp & 1 & Y & (Ta16;Ta16;K22B;Ri17) \\
27 & ESO~113-10 & 0.0257 & 112.2 & 1.8 & 3.1 & 42.68 & 6.74 & Xvar & 2 & n & (NED;VC10;Po12;Bi09) \\
28 & ESO~323-77 & 0.0150 & 65.0 & 1.2 & 9.2 & 42.88 & 6.52 & Ha & 8 & n & (To10;VC10;K22A;Ri17) \\
29 & ESO~33-2 & 0.0181 & 78.6 & 2.0 & 1.4 & 42.12 & 7.60 & Vdisp & 1 & n & (Wa21;VC10;K22B;Ri17) \\
30 & GRS~1734-292 & 0.0218 & 94.9 & 1.0 & 51.9 & 43.89 & 7.84 & Ha & 1 & n & (K22A;VC10;K22A;Ri17) \\
31 & I~Zw~1 & 0.0611 & 273.8 & NLS1 & 4.9 & 43.65 & 6.97 & RM & 3 & Y & (Mi19;Mi19;Hu19;Bi09) \\
32 & IC~4329A & 0.0160 & 69.4 & 1.2 & 96.3 & 43.83 & 7.78 & RM & 6 & Y & (La14;La14;Be23;Ri17) \\
33 & IC~5063 & 0.0110 & 45.9 & 2.0 & 8.3 & 43.02 & 7.74 & Vdisp & 1 & Y & (Mi19;Mi19;Wo02;Ri17) \\
34 & III~Zw~2 & 0.0890 & 367.6 & 1.2 & 7.7 & 44.09 & 8.07 & RM & 1(*) & n & (Me19;Me19;Be15;Ri17) \\
35 & IRAS~F00183-7111 & 0.3270 & 1715.6 & 2.0 & 0.01 & 44.32 & 8.66 & Ms & 1 & Y & (Iw17;VC10;RV;Iw17) \\
36 & IRAS~00521-7054 & 0.0689 & 310.5 & 2.0 & 2.4 & 43.40 & 7.70 & others & 1 & Y & (Wa19;Wa19;Wa19;Ri14) \\
37 & IRAS~04416+1215 & 0.0889 & 406.1 & 1.0 & 1.1 & 43.41 & 6.78 & RM & 3(*) & Y & (Tr22;Tr22;Du15;Tr22) \\
38 & IRAS~05054+1718(W) & 0.0175 & 75.96 & 1.9 & 9.8 & 42.84 & 7.27 & Ms & 2 & Y & (Ya21;Bo23;Ya21;Ya21) \\
39 & IRAS~05078+1626 & 0.0170 & 73.8 & 1.5 & 24.0 & 43.47 & 6.88 & RM & 1 & n & (La14;La14;Be15;Ri17) \\
40 & IRAS~05189-2524 & 0.0426 & 188.4 & 2.0 & 2.9 & 43.40 & 7.40 & Vdisp & 1 & Y & (Sm19;Sm19;K22A;Ri17) \\
41 & IRAS~09149-6206 & 0.0573 & 256.1 & 1.0 & 8.5 & 43.97 & 8.63 & Hb & 6 & n & (Wa20;VC10;K22A;Ri17) \\
42 & IRAS~11119+3257 & 0.1890 & 920.2 & NLS1 & 155.9 & 44.24 & 8.00 & Ha & 2 & Y & (Mi19;Pn19;Pn19;Te10) \\
43 & IRAS~13197-1627 & 0.0165 & 71.6 & 1.8 & 2.8 & 43.41 & 8.37 & Vdisp & 5 & n & (Wa18;Wa18;K22A;Ri17) \\
44 & IRAS~13224-3809 & 0.0658 & 295.8 & NLS1 & 0.5 & 42.75 & 6.82 & Hb & 10 & Y & (Pi18;Pi18;Bi09;Bi09) \\
45 & IRAS~13349+2438 & 0.1085 & 502.3 & NLS1 & 2.2 & 43.87 & 8.62 & RM & 9 & Y & (Pa18;Lo03;Bi09;Bi09) \\
46 & IRAS~17020+4544 & 0.0604 & 270.5 & NLS1 & 5.6 & 43.70 & 6.77 & Hb & 10 & n & (Sa18;Sa18;Bi09;Bi09) \\
47 & IRAS~18325-5926 & 0.0198 & 86.1 & 2.0 & 24.5 & 43.37 & 7.76 & Vdisp & 6 & Y & (Mo11;Mo11;K22B;Ri17) \\
48 & IRAS~23226-3843 & 0.0359 & 158.0 & 1.0 & 2.5 & 42.87 & 6.95 & others & 2 & Y? & (Ko23;VC10;Wa07;Ri17) \\
49 & LBQS~1338-0038 & 0.2375 & 1189.0 & 1.0 & 2.0 & 44.52 & 7.74 & Hb & 1 & Y & (Ma23;VC10;Sh11;Ma23) \\
50 & MCG-01-24-12 & 0.0196 & 85.2 & 2.0 & 12.9 & 43.24 & 7.66 & Vdisp & 1 & Y & (Md21;Md21;K22A;Ri17) \\
51 & MCG-03-58-007 & 0.0323 & 141.8 & 2.0 & 3.7 & 43.75 & 8.00 & Vdisp & 18 & Y & (Br22;Br22;Br18;Br22) \\
52 & MCG-5-23-16 & 0.0085 & 36.2 & 1.9 & 87.5 & 43.19 & 7.65 & Vdisp & 2 & Y & (To10;Br07;K22B;Ri17) \\
53 & MCG-6-30-15 & 0.0070 & 30.4 & 1.5 & 36.8 & 42.66 & 6.30 & RM & 10 & n & (La14;La14;Be15;Ri17) \\
54 & MR~2251-178 & 0.0630 & 282.7 & 1.0 & 37.6 & 44.58 & 8.19 & Hb & 11 & Y & (La14;La14;K22A;Ri17) \\
55 & Mrk~1040(=NGC~931) & 0.0167 & 72.4 & 1.0 & 28.5 & 43.41 & 7.41 & Hb & 1 & n & (Re17;VC10;K22A;Ri17) \\
56 & Mrk~1044 & 0.0160 & 69.4 & NLS1 & 5.0 & 42.46 & 6.45 & RM & 8 & n & (Kr21;Kr21;Du15;Ri17) \\
57 & Mrk~1048(=NGC~985) & 0.0430 & 190.2 & 1.0 & 13.8 & 43.78 & 7.33 & RM & 19 & n & (Eb21;Eb21;U22;Ri17) \\
58 & Mrk~205 & 0.0708 & 319.4 & 1.0 & 6.8 & 43.92 & 8.57 & Hb & 1 & Y & (To10;VC10;K22A;Ri17) \\
59 & Mrk~231 & 0.0422 & 186.6 & 1.0 & 0.9 & 42.65 & 7.87 & Rd & 3 & Y & (Mi19;Mi19;Gr23;Ya21) \\
60 & Mrk~273 & 0.0378 & 166.6 & 2.0 & 0.7 & 43.07 & 8.35 & Vdisp & 1 & Y & (Mi19;Mi19;Ya21;Ya21) \\
61 & Mrk~279 & 0.0305 & 133.7 & 1.0 & 12.3 & 43.41 & 7.43 & RM & 6 & Y & (To10;VC10;Be15;Ri17) \\
62 & Mrk~290 & 0.0296 & 129.6 & 1.5 & 7.4 & 43.18 & 7.28 & RM & 5 & Y & (To10;VC10;Be15;Ri17) \\
63 & Mrk~335 & 0.0250 & 109.1 & NLS1 & 6.0 & 43.21 & 7.23 & RM & 10 & Y & (Ga19;Ga19;Be15;Ri17) \\
64 & Mrk~509 & 0.0340 & 149.4 & 1.2 & 35.3 & 44.07 & 8.05 & RM & 15 & Y & (La14;La14;Be15;Ri17) \\
65 & Mrk~590 & 0.0260 & 113.6 & 1.0 & 3.2 & 42.69 & 7.57 & RM & 6 & n & (Gu15;Gu15;Be15;Ri17) \\
66 & Mrk~6 & 0.0190 & 82.6 & 1.5 & 11.9 & 43.10 & 8.10 & RM & 1(*) & n & (Me19;Me19;Be15;Ri17) \\
67 & Mrk~704 & 0.0290 & 127.0 & 1.2 & 10.7 & 43.32 & 7.56 & RM & 3 & n & (La14;La14;Be15;Ri17) \\
68 & Mrk~766 & 0.0129 & 55.8 & NLS1 & 13.2 & 42.71 & 6.82 & RM & 11 & Y & (La14;Ni09;Be15;Ri17) \\
69 & Mrk~79(=UGC~3973) & 0.0222 & 96.7 & 1.2 & 7.7 & 43.11 & 7.61 & RM & 2 & Y & (To10;VC10;Be15;Ri17) \\
70 & Mrk~841 & 0.0364 & 160.2 & 1.5 & 9.7 & 43.67 & 7.66 & RM & 5 & Y & (To10;VC10;U22;Ri17) \\
71 & Mrk~896 & 0.0270 & 118.0 & NLS1 & 3.5 & 42.79 & 6.35 & Hb & 1(*) & n & (Me19;Pg03;Bi09;Bi09) \\
72 & NGC~1068 & 0.0038 & 14.4 & 2.0 & 4.9 & 43.04 & 7.23 & maser & 1(*) & Y & (Mi19;Mi19;Ga23;Ri17) \\
73 & NGC~1365 & 0.0055 & 19.6 & 1.8 & 17.4 & 42.38 & 6.26 & Rd & 10 & n & (Br14;Br14;Gr23;Ri17) \\
74 & NGC~1566 & 0.0050 & 17.9 & 1.5 & 3.1 & 41.56 & 6.92 & Vdisp & 2 & n & (Pa19;VC10;Wo02;Ri17) \\
75 & NGC~2992 & 0.0077 & 38.0 & 1.9 & 7.3 & 42.16 & 7.48 & Hb & 5 & Y & (Lu23;VC10;Gu21;Ri17) \\
76 & NGC~3227 & 0.0030 & 22.9 & 1.5 & 36.9 & 42.28 & 6.68 & RM & 8 & n & (La14;La14;Be15;Ri17) \\
77 & NGC~3516 & 0.0080 & 38.9 & 1.5 & 27.5 & 42.46 & 7.40 & RM & 9 & n & (La14;La14;Be15;Ri17) \\
78 & NGC~3783 & 0.0090 & 38.5 & 1.0 & 76.9 & 43.24 & 7.08 & RM & 20 & n & (La14;La14;Be15;Ri17) \\
79 & NGC~4051 & 0.0020 & 11.0 & 1.5 & 17.6 & 41.07 & 5.89 & RM & 37 & Y & (La14;La14;Be15;Ri17) \\
80 & NGC~4151 & 0.0033 & 19.0 & 1.5 & 47.2 & 42.66 & 7.37 & RM & 5 & Y & (To10;VC10;Be15;Ri17) \\
81 & NGC~4388 & 0.0084 & 18.1 & 2.0 & 20.4 & 42.92 & 6.92 & maser & 2 & n & (Sh08;Sh08;Ku11;Ri17) \\
82 & NGC~4395 & 0.0011 & 4.8 & 1.8 & 5.5 & 40.56 & 5.45 & RM & 1 & n & (Na23;VC10;Be15;Ri17) \\
83 & NGC~4507 & 0.0118 & 51.0 & 2.0 & 7.4 & 43.51 & 7.81 & Vdisp & 1 & Y & (To10;Mt04;K22B;Ri17) \\
84 & NGC~4593 & 0.0090 & 37.2 & 1.0 & 35.7 & 42.91 & 6.91 & RM & 11 & n & (La14;La14;Be15;Ri17) \\
85 & NGC~5506 & 0.0062 & 26.4 & 1.9 & 92.7 & 43.08 & 7.24 & Vdisp & 1 & Y & (Bl05;Bl05;K22B;Ri17) \\
86 & NGC~5548 & 0.0170 & 73.8 & 1.5 & 19.1 & 43.13 & 7.69 & RM & 19 & n & (La14;La14;Be15;Ri17) \\
87 & NGC~6240 & 0.0245 & 106.9 & 2.0 & 2.3 & 43.86 & 8.53 & Vdisp & 2(*) & Y & (Mi19;Mi19;En10;Ya21) \\
88 & NGC~6860 & 0.0149 & 64.5 & 1.5 & 23.3 & 43.11 & 7.71 & Vdisp & 2 & n & (Wi10;Wi10;K22B;Ri17) \\
89 & NGC~7469 & 0.0160 & 69.4 & 1.2 & 25.4 & 43.17 & 6.96 & RM & 13 & n & (La14;La14;Be15;Ri17) \\
90 & NGC~7582 & 0.0053 & 22.5 & 2.0 & 4.0 & 41.74 & 7.08 & Vdisp & 1 & Y & (To10;Bi09;K22B;Rv15) \\
91 & PDS~456 & 0.1840 & 893.2 & 1(QSO) & 7.0 & 44.90 & 8.23 & VLTI & 28 & Y & (Na15;Re09;GR24;Bi09) \\
92 & PG~0804+761 & 0.1000 & 460.3 & 1.0 & 10.8 & 44.45 & 8.73 & RM & 1 & Y & (Ma23;VC10;Be15;Ri17) \\
93 & PG~0844+349 & 0.0640 & 287.4 & 1.0 & 4.8 & 43.69 & 7.86 & RM & 1(*) & Y & (Po03;VC10;Be15;Bi09) \\
94 & PG~0947+396 & 0.2055 & 1010.3 & 1.0 & 1.7 & 44.33 & 8.21 & RM & 1 & Y & (Ma23;VC10;Wo24;Bi09) \\
95 & PG~1114+445 & 0.1440 & 682.1 & 1.0 & 2.2 & 44.09 & 8.59 & Hb & 11 & Y & (Ma23;VC10;Bi09;Bi09) \\
96 & PG~1126-041 & 0.0600 & 268.7 & 1.0 & 1.2 & 43.25 & 8.10 & Vdisp & 8 & Y & (Gi11;VC10;Gi11;Gi11) \\
97 & PG~1202+281 & 0.1650 & 791.8 & 1.2 & 5.2 & 44.59 & 8.06 & RM & 1(*) & Y & (Ma23;VC10;Wo24;Ri17) \\
98 & PG~1211+143 & 0.0809 & 367.6 & NLS1 & 3.0 & 43.73 & 8.16 & RM & 17 & Y & (To10;Ni09;Pe04;Bi09) \\
99 & PG~1402+261 & 0.1640 & 786.5 & NLS1 & 1.7 & 44.13 & 7.53 & RM & 1 & Y & (Re04;Re04;Hu21;Bi09) \\
100 & PG~1404+226 & 0.0980 & 450.5 & NLS1 & 0.1 & 42.48 & 6.83 & RM & 1 & n & (Da05;Da05;Hu21;Bi09) \\
101 & PG~1448+273 & 0.0645 & 289.7 & NLS1 & 1.9 & 43.31 & 7.00 & RM & 6 & Y & (Ko20;Ko20;Hu21;Bi09) \\
102 & PKS~0405-12 & 0.5740 & 3345.6 & 1.2 & 4.3 & 44.37 & 9.47 & others & 1(*) & n & (Me19;Me19;Wo02;Me19) \\
103 & PKS~0921-213 & 0.0530 & 236.1 & 1.0 & 5.6 & 43.57 & 8.40 & Hb & 1(*) & n & (Me19;Me19;K22A;Ri17) \\
104 & PKS~1549-79 & 0.1522 & 724.6 & NLS1 & 5.7 & 44.72 & 8.43 & Vdisp & 2 & Y & (To14;Ho06;K22B;Ri17) \\
105 & PKS~2135-14 & 0.2000 & 980.1 & 1.5 & 5.4 & 44.78 & 9.10 & Hb & 1(*) & n & (Me19;Me19;K22A;Ri17) \\
106 & SWIFT~J2127.4+5654 & 0.0147 & 63.7 & NLS1 & 23.5 & 43.09 & 7.15 & Hb & 1 & Y & (Sa13;Sa13;K22A;Ri17) \\
107 & Ton~28 & 0.3290 & 1727.8 & 1.0 & 0.5 & 44.23 & 7.57 & RM & 1 & Y & (Na19;VC10;Be15;Na19) \\
108 & Ton~S180 & 0.0620 & 278.0 & NLS1 & 4.7 & 43.65 & 7.06 & RM & 1 & n & (Ma20;Ma20;Bi09;Bi09) \\
109 & WKK~4438 & 0.0160 & 69.4 & NLS1 & 9.2 & 42.74 & 6.30 & Hb & 1 & Y & (Ji18;Ji18;Ji18;Ri17) \\
\hline
\multicolumn{12}{c}{High-$z$ (z$>$1) Galaxies} \\
\hline
110 & APM~08279+5255 & 3.91 & 34896 & QSO & 0.4 & 46.19 & 9.87 & RM & 12 & Y & (Ch21;Ch21;Be15;Ch21) \\
111 & CXOCDFS~J0332-2747 & 2.579 & 21180 & QSO & 0.01 & 44.66 & 9.41 & Ms & 1 & Y? & (Wn05;Wn05;RC;To06) \\
112 & H~1413+117 & 2.56 & 20992 & QSO & 0.04 & 43.87 & 8.60 & CIV & 2 & Y? & (Ch07;Ch07;OD15;Ru21) \\
113 & HS~0810+2554 & 1.51 & 11000 & QSO & 0.4 & 43.73 & 8.60 & Hb & 5 & Y & (Ch21;Ch21;Ch21;Ch21) \\
114 & HS~1700+6416 & 2.735 & 22740 & QSO & 0.1 & 45.34 & 10.20 & CIV & 4 & Y & (Ch21;Ch21;Ch21;Ch21) \\
115 & MG~J0414+0534 & 2.64 & 21789 & QSO & 0.3 & 44.49 & 9.00 & Hb & 1 & Y & (Ch21;Ch21;Ch21;Ch21) \\
116 & PG~1115+080 & 1.72 & 12914 & QSO & 0.4 & 44.15 & 8.80 & Hb & 2 & Y & (Ch21;Ch21;Ch21;Ch21) \\
117 & PG~1247+268 & 2.048 & 15997 & QSO & 0.3 & 45.91 & 9.80 & CIV & 1(*) & Y? & (La16;La16;Sh11;Bi09) \\
118 & PID352 & 1.6 & 11814 & QSO & 0.03 & 45.43 & 8.70 & Ms & 1 & Y & (Ch21;Ch21;Vi15;Ch21) \\
119 & Q~2237+0305 & 1.695 & 12683 & QSO & 0.2 & 44.12 & 9.10 & Hb & 2 & Y & (Ch21;Ch21;Ch21;Ch21) \\
120 & SDSS~J0904+1512 & 1.826 & 13899 & QSO & 0.1 & 44.23 & 9.30 & CIV & 1 & Y & (Ch21;Ch21;Ch21;Ch21) \\
121 & SDSS~J0921+2854 & 1.41 & 10107 & QSO & 0.7 & 45.21 & 8.90 & CIV & 2 & Y & (Ch21;Ch21;Ch21;Ch21) \\
122 & SDSS~J1029+2623 & 2.197 & 17430 & QSO & 0.2 & 44.04 & 8.80 & CIV & 1 & Y & (Ch21;Ch21;Ch21;Ch21) \\
123 & SDSS~J1128+2402 & 1.608 & 11886 & QSO & 0.1 & 44.32 & 8.70 & CIV & 1 & Y & (Ch21;Ch21;Ch21;Ch21) \\
124 & SDSS~J1353+1138 & 1.627 & 12060 & QSO & 0.1 & 44.70 & 9.40 & CIV & 1 & Y & (Ch21;Ch21;Ch21;Ch21) \\
125 & SDSS~J1442+4055 & 2.593 & 21320 & QSO & 0.3 & 44.75 & 9.70 & CIV & 1 & Y & (Ch21;Ch21;Ch21;Ch21) \\
126 & SDSS~J1529+1038 & 1.984 & 15388 & QSO & 0.03 & 44.04 & 8.90 & CIV & 1 & Y & (Ch21;Ch21;Ch21;Ch21) \\
\hline
\multicolumn{12}{c}{X-WING Candidates} \\
\hline
127 & MCG-02-14-009 & 0.0285 & 124.7 & 1.0 & 4.2 & 42.90 & 7.13 & Xvar & 1 & Y? & (Ka16;Ka16;Po12;Ri17) \\
128 & NGC~7213 & 0.0058 & 22.0 & 1.5 & 11.8 & 41.95 & 7.13 & Hb & 1 & Y? & (La14;La14;K22A;Ri17) \\
129 & NGC~7314 & 0.0048 & 16.8 & 1.9 & 38.4 & 42.24 & 6.30 & Vdisp & 1 & Y? & (Zo13;Zo13;K22B;Ri17) \\
130 & PG~1244+026 & 0.0481 & 213.5 & NLS1 & 1.9 & 43.07 & 6.52 & Hb & 1 & Y? & (SIM;Al14;Bi09;Bi09) \\
131 & PKS~0558-504 & 0.1370 & 646.0 & NLS1 & 13.6 & 44.93 & 7.62 & Hb & 1 & Y? & (Gh16;Gh16;K22A;Ri17) \\
132 & RE~J1034+396 & 0.0420 & 185.6 & NLS1 & 0.6 & 42.53 & 6.81 & Hb & 1 & Y? & (Jn20;Jn20;Bi09;Bi09) \\
\enddata
\tablecomments{Columns:
(1)--(2) ID and object name;
(3) redshift;
(4) Light distance: For the closest objects at $D_{\rm L} < 50$~Mpc, we adopted redshift-independent distance measurements \citep{Koss2022b}.
(5) Optical Classes
(6)--(7) observed 2--10~keV fluxes ($F_{\rm 2-10}$) in units of $10^{-12}$~erg~cm$^{-2}$~s$^{-1}$ and logarithmic intrinsic AGN luminosity in the 2--10~keV band in units of erg~s$^{-1}$, many of which are referred from \citet{Ricci2017d} and \citet{Bianchi2009a};
(8) as logarithmic $M_{\rm BH}$ in units of $M_{\odot}$.
(9) Methods of $M_{\rm BH}$ estimation: 
reverberation mapping (RM), 
H\textsubscript{2}O megamesar (maser),
broad emission lines of H$\alpha$ (Ha), H$\beta$ (Hb), and \ion{C}{4} (CIV), 
near-infrared interferometric observations of the BLR with Very Large Telescope Interferometer (VLTI);
stellar velocity dispersion (Vdisp), 
stellar mass to $M_{\rm BH}$ ratio (Ms), 
X-ray variability (Xvar), 
and others (others);
(10) Number of X-ray winds reported in the study.
The symbol (*) marks the uncertain AGNs with X-ray winds as explained in Section~\ref{sub2-2-winds}.
(11) Flag for the presence of UFOs with \ion{Fe}{25}/\ion{Fe}{26} lines detected in the $>$6~keV band (Section~\ref{sub3-3-Fe}).
``Y'' means the AGNs with UFOs.
``Y?'' denotes that the AGNs with UFO candidates, all of which only $V_{\rm out}$ were constrained (but not $\xi$ and $N_{\rm H}$; see also Section~\ref{sub2-2-winds}).
``n'' indicates AGNs with warm absorbers but no UFOs;
(12) References for the data in columns (3)--(4), (5), (6)--(7), and (8)--(9). \\
References:
(Al14) \citet{Alston2014a};
(Be15) \citet{Bentz2015};
(Be23) \citet{Bentz2023a};
\ifnum 0=0
(Bi09) \citet{Bianchi2009a};
(Bl05) \citet{Blustin2005};
(Bo23) \citet{Bonanomi2023};
(Br07) \citet{Braito2007};
(Br14) \citet{Braito2014};
(Br18) \citet{Braito2018};
(Br22) \citet{Braito2022};
(Ch07) \citet{Chartas2007b};
(Ch19) \citet{Cheng2019};
(Ch21) \citet{Chartas2021};
(Da05) \citet{Dasgupta2005a};
(Du15) \citet{Du2015};
(Eb21) \citet{Ebrero2021};
(En10) \citet{Engel2010};
(Ga13) \citet{Gallo2013b};
(Ga19) \citet{Gallo2019b};
(Ga23) \citet{Gallimore2023};
(Gh16) \citet{Ghosh2016a};
(Gi11) \citet{Giustini2011};
(GR24) \citet{GRAVITY2024};
(Gu15) \citet{Gupta2015};
(Gu21) \citet{Guolo2021};
(Ho06) \citet{Holt2006d};
(Hu19) \citet{Huang2019};
(Hu21) \citet{Hu2021};
(Iw17) \citet{Iwasawa2017};
(Ji18) \citet{Jiang2018a};
(Ji18) \citet{Jiang2018b};
(Jn20) \citet{Jin2020};
(K22A) \citet{Koss2022b};
(K22B) \citet{Koss2022c};
(Ka16) \citet{Kara2016};
(Ka21) \citet{Kara2021};
(Ko20) \citet{Kosec2020b};
(Ko23) \citet{Kollatschny2023};
(Kr21) \citet{Krongold2021};
(Ku11) \citet{Kuo2011};
(La14) \citet{Laha2014b};
(La16) \citet{Lanzuisi2016};
(Lo03) \citet{Longinotti2003};
(Lu23) \citet{Luminari2023a};
(Ma20) \citet{Matzeu2020};
(Ma23) \citet{Matzeu2023};
(Md20) \citet{Middei2020};
(Md21) \citet{Middei2021};
(Me19) \citet{Mehdipour2019};
(Mi19) \citet{Mizumoto2019a};
(Mo11) \citet{Mocz2011};
(Mt04) \citet{Matt2004b};
(NED) NASA/IPAC Extragalactic Database;
(Na15) \citet{Nardini2015a};
(Na19) \citet{Nardini2019a};
(Na23) \citet{Nandi2023};
(Ni09) \citet{Nikolajuk2009};
(OD15) \citet{ODowd2015};
(Pa18) \citet{Parker2018d};
(Pa19) \citet{Parker2019a};
(Pe04) \citet{Peterson2004};
(Pg03) \citet{Page2003c};
(Pi18) \citet{Pinto2018a};
(Pl14) \citet{Paliya2014};
(Pn19) \citet{Pan2019b};
(Po03) \citet{Pounds2003d};
(Po07) \citet{Pounds2007b};
(Po12) \citet{Ponti2012};
(RC) \citet{Reines2015} and \citet{Corral2016};
(RV) \citet{Reines2015} and \citet{Vika2017};
(Re04) \citet{Reeves2004b};
(Re09) \citet{Reeves2009a};
(Re17) \citet{Reeves2017};
(Ri14) \citet{Ricci2014c};
(Ri17) \citet{Ricci2017d};
(Ri20) \citet{Ricci2020};
(Ru21) \citet{Ruiz2021};
(Rv15) \citet{Rivers2015b};
(SIM) Simbad;
(Sa13) \citet{Sanfrutos2013};
(Sa18) \citet{Sanfrutos2018};
(Se07) \citet{Sergeev2007};
(Sh08) \citet{Shirai2008};
(Sh11) \citet{Shen2011};
(Si07) \citet{Sikora2007};
(Sm19) \citet{Smith2019};
(Ta16) \citet{Tatum2016};
(Te10) \citet{Teng2010};
(To06) \citet{Tozzi2006};
(To10) \citet{Tombesi2010c};
(To14) \citet{Tombesi2014a};
(Tr22) \citet{Tortosa2022};
(U22)  \citet{U2022};
(VC10) \citet{Veron-Cetty2010};
(Vi15) \citet{Vignali2015};
(Wa07) \citet{Wang2007};
(Wa18) \citet{Walton2018};
(Wa19) \citet{Walton2019};
(Wa20) \citet{Walton2020b};
(Wa21) \citet{Walton2021};
(Wi10) \citet{Winter2010c};
(Wn05) \citet{Wang2005};
(Wo02) \citet{Woo2002};
(Wo24) \citet{Woo2024};
(Xu21) \citet{XuYerong2021};
(Xu22) \citet{XuYerong2022};
(Ya21) \citet{Yamada2021};
(Zo13) \citet{Zoghbi2013b}.
\fi
}
\tablenotetext{\ }{(This table is available in its entirety in machine-readable form.)}
\end{deluxetable*}

\clearpage

\startlongtable
\begin{deluxetable*}{llccccccccc}
\label{TB3-outflow}
\tablecaption{X-ray Wind Properties}
\tabletypesize{\footnotesize}
\tablehead{
\colhead{ID} &
\colhead{Object} &
\colhead{Start date} &
\colhead{Inst.} &
\colhead{Fe-K} &
\colhead{log$N_{\rm H}$} &
\colhead{log$V_{\rm out}$} &
\colhead{log$\xi$} &
\colhead{log$L_{\rm ion}$} &
\colhead{Type} &
\colhead{Model}
}
\decimalcolnumbers
\startdata
1a & 1E~0754.6+3928 & 2006-04-18 & P & Y & $23.11^{+0.25}_{-0.41}$ & $4.84^{+0.04}_{-0.04}$ & $3.40^{+0.30}_{-0.30}$ & 44.11 & UFO & XSTAR \\
1b & 1E~0754.6+3928 & 2006-04-18 & P & Y & $23.41^{+0.27}_{-0.51}$ & $3.71^{+0.09}_{-0.12}$ & $3.40^{+0.10}_{-0.10}$ & 44.11 & WA & XSTAR \\
1c & 1E~0754.6+3928 & 2006-04-18 & P & n & $22.85^{+0.06}_{-0.07}$ & $<$3.18 & $2.00^{+0.05}_{-0.05}$ & 44.11 & WA & XSTAR \\
1d & 1E~0754.6+3928 & 2006-10-22 & P & n & $22.60^{+0.05}_{-0.06}$ & $<$3.04 & $1.50^{+0.07}_{-0.07}$ & 44.11 & WA & XSTAR \\
2a & 1ES~1927+654 & 2011-04-16to2011-05-20 & PSu & n & $23.82^{+0.09}_{-0.05}$ & $4.90^{+0.01}_{-0.01}$ & $3.03^{+0.11}_{-0.16}$ & (non) & UFO & zxipcf \\
2b & 1ES~1927+654 & 2011-04-16to2011-05-20 & PSu & n & $22.52^{+0.24}_{-0.15}$ & $4.90^{+0.01}_{-0.01}$ & $-0.30^{+0.01}_{-0.01}$ & (non) & LIP & zxipcf \\
3a & 1H~0323+342 & 2015-08-23 & RPM & n & $20.86^{+0.07}_{-0.08}$ & $2.94^{+0.09}_{-0.11}$ & $0.15^{+0.20}_{-0.20}$ & (non) & WA & xabs \\
3b & 1H~0323+342 & 2015-08-23 & RPM & n & $20.95^{+0.14}_{-0.20}$ & $2.92^{+0.13}_{-0.18}$ & $2.17^{+0.03}_{-0.03}$ & (non) & WA & xabs \\
4a & 1H~0419-577 & 2002-09-25 & P & Y & $<$23.61 & $4.37^{+0.06}_{-0.07}$ & $3.69^{+0.87}_{-0.87}$ & 44.60 & UFO & XSTAR \\
5a & 1H~0707-495 & 2000-10-21to2019-10-11 & RP & Y & $22.26^{+0.16}_{-0.26}$ & $4.73^{+0.02}_{-0.03}$ & $3.93^{+0.08}_{-0.10}$ & 44.15 & UFO & xabs \\
\enddata
\tablecomments{Columns:
(1) Detection ID of outflow components
(2) Object names
(3) Start date of observation. The start dates of the oldest and latest observations are shown when multiple observations are used to identify the ionized outflows. 
(4) Instruments of observation (A=Chandra/ACIS, F=NuSTAR/FPM, H=Chandra/HETGS, L=Chandra/LETGS, M=XMM-Newton/MOS, P=XMM-Newton/PN, R=XMM-Newton/RGS, Su=Suzaku/XIS, and Sw=Swift/XRT).
(5) The method of outflow detection utilizes blueshifted Fe-K absorption lines, \ion{Fe}{25}/\ion{Fe}{26} K$\alpha$/K$\beta$ above the rest-frame 6~keV band (Y) or absorption lines below the 6~keV band (n).
(6)--(8) Logarithmic $N_{\rm H}$ of the ionized outflows in units of cm$^{-2}$, $V_{\rm out}$ in units of km~s$^{-1}$, and $\xi$ in units of erg~s$^{-1}$~cm.
(*) indicates that the values are presented assuming ${\log}U = {\log}\xi - 1.75$ (Section~\ref{sub2-2-winds}).
When the uncertainties of $V_{\rm out}$ or $N_{\rm H}$ are not constrained, we assume that they are $\pm$0.5, which are conservative values larger than the typical errors ($\sim$0.1--0.2) in Figures~\ref{F3-outflow-opt-1}--\ref{F7-outflow-radius}.
(9) Logarithmic ionizing luminosity in units of erg~s$^{-1}$.
(10) Outflow-type UFO, low-$\xi$ UFO (LIP) and warm absorber (WA).
The OBSIDs and the references for each ID are listed in Table~\ref{TB4-obsid}.
(11) The types of the X-ray fitting models of the X-ray winds (see Appendix~\ref{AppendixA-inst}).
}
\tablenotetext{\ }{(This table is available in its entirety in machine-readable form.)}
\end{deluxetable*}


\startlongtable
\begin{deluxetable*}{clllllllllll}
\label{TB4-obsid}
\tablecaption{Observations and References of the X-ray Winds}
\tabletypesize{\footnotesize}
\tablehead{
\colhead{ID} &
\colhead{Object} &
\colhead{OBSID} &
\colhead{Ref1} &
\colhead{Ref2} &
\colhead{Ref3} &
\colhead{Ref4} &
\colhead{Ref5} &
\colhead{Ref6} &
\colhead{Ref7} &
\colhead{Ref8} &
\colhead{Ref9}
}
\decimalcolnumbers
\startdata
1a & 1E~0754.6+3928 & 305990101 & Mid20 &  &  &  &  &  &  &  &  \\
1b & 1E~0754.6+3928 & 305990101 & Mid20 &  &  &  &  &  &  &  &  \\
1c & 1E~0754.6+3928 & 305990101 & Mid20 &  &  &  &  &  &  &  &  \\
1d & 1E~0754.6+3928 & 406740101 & Mid20 &  &  &  &  &  &  &  &  \\
2a & 1ES~1927+654 & 0671860201+706006010 & Gal13b &  &  &  &  &  &  &  &  \\
2b & 1ES~1927+654 & 0671860201+706006010 & Gal13b &  &  &  &  &  &  &  &  \\
3a & 1H~0323+342 & 764670101 & Meh19 &  &  &  &  &  &  &  &  \\
3b & 1H~0323+342 & 764670101 & Meh19 &  &  &  &  &  &  &  &  \\
4a & 1H~0419-577 & 148000201 & Tom12a & Tom11a & DiG14 &  &  &  &  &  &  \\
5a & 1H~0707-495 & 16 OBSIDs & XuY21 & Par21 & Kos18 & Hag16 & Dau12 & Gal04 &  &  &  \\
\enddata
\tablecomments{Columns:
(1) Detection ID of outflow components
(2) Object names
(3) OBSIDs;
(4) References for the adopted X-ray winds
(5)--(12) pertinent papers on duplicated reports.\\
References: 
(And10) \citet{Andrade-Velazquez2010};
(Bal11) \citet{Ballo2011b};
(Bal15) \citet{Ballo2015};
\ifnum 0=0
(Beh03) \citet{Behar2003};
(Beh10) \citet{Behar2010};
(Beh17) \citet{Behar2017};
(Ber20) \citet{Bertola2020};
(Beu15) \citet{Beuchert2015};
(Beu17) \citet{Beuchert2017};
(Blu02) \citet{Blustin2002};
(Blu03) \citet{Blustin2003};
(Blu05) \citet{Blustin2005};
(Blu07) \citet{Blustin2007};
(Blu09) \citet{Blustin2009};
(Boi19) \citet{Boissay-Malaquin2019};
(Bon23) \citet{Bonanomi2023};
(Bra07) \citet{Braito2007};
(Bra11) \citet{Braito2011};
(Bra14) \citet{Braito2014};
(Bra18) \citet{Braito2018};
(Bra21) \citet{Braito2021};
(Bra22) \citet{Braito2022};
(Bre12) \citet{Brenneman2012};
(Bre13) \citet{Brenneman2013b};
(Bri06) \citet{Brinkmann2006};
(Cap09) \citet{Cappi2009};
(Cap16) \citet{Cappi2016};
(Cha02) \citet{Chartas2002};
(Cha03) \citet{Chartas2003};
(Cha07a) \citet{Chartas2007a};
(Cha07b) \citet{Chartas2007b};
(Cha09) \citet{Chartas2009};
(Cha14) \citet{Chartas2014};
(Cha16) \citet{Chartas2016};
(Cha18) \citet{Chartas2018};
(Cha21) \citet{Chartas2021};
(Chi11) \citet{Chiang2011};
(Chi12) \citet{Chiang2012};
(Col01) \citet{Collinge2001};
(Cos07a) \citet{Costantini2007a};
(Cos07b) \citet{Costantini2007b};
(Cos10) \citet{Costantini2010a};
(Dad05) \citet{Dadina2005};
(Dad18) \citet{Dadina2018};
(Dan18) \citet{Danehkar2018c};
(Das05) \citet{Dasgupta2005a};
(Dau12) \citet{Dauser2012};
(Det08) \citet{Detmers2008};
(Det10) \citet{Detmers2010};
(Det11) \citet{Detmers2011};
(DiG14) \citet{DiGesu2014};
(DiG15) \citet{DiGesu2015};
(DiG16) \citet{DiGesu2016};
(Ebr10) \citet{Ebrero2010};
(Ebr11a) \citet{Ebrero2011a};
(Ebr11b) \citet{Ebrero2011b};
(Ebr13) \citet{Ebrero2013};
(Ebr16a) \citet{Ebrero2016a};
(Ebr16b) \citet{Ebrero2016b};
(Ebr21) \citet{Ebrero2021};
(Fer15) \citet{Feruglio2015};
(Gal04) \citet{Gallo2004};
(Gal13b) \citet{Gallo2013b};
(Gal13c) \citet{Gallo2013c};
(Gal15) \citet{Gallo2015};
(Gal19) \citet{Gallo2019b};
(Gan22) \citet{Gandhi2022b};
(Gib05) \citet{Gibson2005a};
(Gib07) \citet{Gibson2007};
(Giu11) \citet{Giustini2011};
(Giu23) \citet{Giustini2023};
(Gof11) \citet{Gofford2011};
(Gof13) \citet{Gofford2013};
(Gof14) \citet{Gofford2014};
(Gof15) \citet{Gofford2015};
(Gup13a) \citet{Gupta2013a};
(Gup13c) \citet{Gupta2013c};
(Gup15) \citet{Gupta2015};
(Hag15) \citet{Hagino2015};
(Hag16) \citet{Hagino2016};
(Hag17) \citet{Hagino2017};
(Hol05) \citet{Holczer2005};
(Hol07) \citet{Holczer2007};
(Hol10) \citet{Holczer2010};
(Hol12) \citet{Holczer2012};
(Igo20) \citet{Igo2020};
(Iwa16) \citet{Iwasawa2016};
(Iwa17) \citet{Iwasawa2017};
(Jia18a) \citet{Jiang2018a};
(Jia18b) \citet{Jiang2018b};
(Jia22) \citet{Jiang2022};
(Jim08) \citet{Jimenez-Bailon2008};
(Kaa02) \citet{Kaastra2002};
(Kaa04) \citet{Kaastra2004};
(Kaa11) \citet{Kaastra2011b};
(Kaa12) \citet{Kaastra2012};
(Kaa14a) \citet{Kaastra2014a};
(Kaa14b) \citet{Kaastra2014b};
(Kar17) \citet{Kara2017};
(Kar21) \citet{Kara2021};
(Kas00) \citet{Kaspi2000};
(Kas01) \citet{Kaspi2001};
(Kas02) \citet{Kaspi2002};
(Kas04) \citet{Kaspi2004};
(Kas06) \citet{Kaspi2006};
(Kin12) \citet{King2012};
(Kol20) \citet{Kollatschny2020};
(Kol23) \citet{Kollatschny2023};
(Kos18) \citet{Kosec2018};
(Kos20) \citet{Kosec2020b};
(Kra05) \citet{Kraemer2005};
(Kro03) \citet{Krongold2003};
(Kro05a) \citet{Krongold2005a};
(Kro05b) \citet{Krongold2005b};
(Kro07) \citet{Krongold2007};
(Kro09) \citet{Krongold2009};
(Kro10) \citet{Krongold2010b};
(Kro21) \citet{Krongold2021};
(Lah11) \citet{Laha2011};
(Lah13) \citet{Laha2013};
(Lah14) \citet{Laha2014b};
(Lah16) \citet{Laha2016b};
(Lan12) \citet{Lanzuisi2012b};
(Lan16) \citet{Lanzuisi2016};
(Lau21) \citet{Laurenti2021};
(Lee13) \citet{Lee2013};
(Lob11) \citet{Lobban2011};
(Lob16) \citet{Lobban2016};
(Lon03) \citet{Longinotti2003};
(Lon10) \citet{Longinotti2010};
(Lon13) \citet{Longinotti2013};
(Lon15) \citet{Longinotti2015};
(Lon19) \citet{Longinotti2019};
(Lum18) \citet{Luminari2018};
(Lum23) \citet{Luminari2023a};
(Mal18) \citet{Mallick2018b};
(Mao17) \citet{Mao2017};
(Mao19) \citet{Mao2019};
(Mao22) \citet{Mao2022};
(Mar06) \citet{Markowitz2006};
(Mar09) \citet{Markowitz2009a};
(Mari18) \citet{Marinucci2018b};
(Mas03) \citet{Mason2003};
(Mat16) \citet{Matzeu2016};
(Mat17) \citet{Matzeu2017};
(Mat19) \citet{Matzeu2019};
(Mat20) \citet{Matzeu2020};
(Mat23) \citet{Matzeu2023};
(Mats04) \citet{Matsumoto2004};
(Matt04) \citet{Matt2004b};
(McK03) \citet{McKernan2003a};
(McK07) \citet{McKernan2007};
(Meh10) \citet{Mehdipour2010};
(Meh12) \citet{Mehdipour2012};
(Meh17) \citet{Mehdipour2017};
(Meh18) \citet{Mehdipour2018a};
(Meh19) \citet{Mehdipour2019};
(Meh23) \citet{Mehdipour2023b};
(Mid20) \citet{Middei2020};
(Mid21) \citet{Middei2021};
(Mid23) \citet{Midooka2023};
(Min07) \citet{Miniutti2007};
(Min14) \citet{Miniutti2014};
(Miz17) \citet{Mizumoto2017};
(Miz19a) \citet{Mizumoto2019a};
(Miz19b) \citet{Mizumoto2019b};
(Miz21) \citet{Mizumoto2021};
(Moc23) \citet{Mochizuki2023};
(Moc11) \citet{Mocz2011};
(Mon22) \citet{Mondal2022};
(Nar14) \citet{Nardini2014};
(Nar15) \citet{Nardini2015a};
(Nar18) \citet{Nardini2018};
(Nar19) \citet{Nardini2019a};
(Net03) \citet{Netzer2003b};
(Ogl04) \citet{Ogle2004};
(Ogo22) \citet{Ogorzalek2022};
(Pag03) \citet{Page2003c};
(Pap07) \citet{Papadakis2007a};
(Pap16) \citet{Papadakis2016b};
(Par17a) \citet{Parker2017a};
(Par17b) \citet{Parker2017b};
(Par18a) \citet{Parker2018a};
(Par18d) \citet{Parker2018d};
(Par19a) \citet{Parker2019a};
(Par19c) \citet{Parker2019c};
(Par20) \citet{Parker2020b};
(Par21) \citet{Parker2021};
(Pat12) \citet{Patrick2012};
(Pin18) \citet{Pinto2018a};
(Pon09) \citet{Ponti2009};
(Pou03c) \citet{Pounds2003c};
(Pou03d) \citet{Pounds2003d};
(Pou04) \citet{Pounds2004b};
(Pou06) \citet{Pounds2006b};
(Pou07a) \citet{Pounds2007a};
(Pou07b) \citet{Pounds2007b};
(Pou09) \citet{Pounds2009};
(Pou11) \citet{Pounds2011a};
(Pou12) \citet{Pounds2012};
(Pou13) \citet{Pounds2013};
(Pou14) \citet{Pounds2014};
(Pou16a) \citet{Pounds2016a};
(Pou16c) \citet{Pounds2016c};
(Pou18) \citet{Pounds2018};
(Ree03) \citet{Reeves2003b};
(Ree04a) \citet{Reeves2004a};
(Ree04b) \citet{Reeves2004b};
(Ree05) \citet{Reeves2005};
(Ree08) \citet{Reeves2008};
(Ree09a) \citet{Reeves2009a};
(Ree09b) \citet{Reeves2009b};
(Ree10) \citet{Reeves2010c};
(Ree13) \citet{Reeves2013};
(Ree14) \citet{Reeves2014};
(Ree16) \citet{Reeves2016a};
(Ree17) \citet{Reeves2017};
(Ree18a) \citet{Reeves2018a};
(Ree18b) \citet{Reeves2018b};
(Ree19) \citet{Reeves2019};
(Ree20) \citet{Reeves2020};
(Ree23) \citet{Reeves2023};
(Ris05) \citet{Risaliti2005c};
(Ris11) \citet{Risaliti2011a};
(Sae09) \citet{Saez2009};
(Sae11) \citet{Saez2011};
(Sak01) \citet{Sako2001};
(Sak03) \citet{Sako2003b};
(San16) \citet{Sanfrutos2016a};
(San18) \citet{Sanfrutos2018};
(Sco14) \citet{Scott2014};
(Ser19) \citet{Serafinelli2019};
(Ser23) \citet{Serafinelli2023};
(Shi08) \citet{Shirai2008};
(Smi07) \citet{Smith2007a};
(Smi08) \citet{Smith2008};
(Smi19) \citet{Smith2019};
(Ste03a) \citet{Steenbrugge2003a};
(Ste03c) \citet{Steenbrugge2003c};
(Ste05a) \citet{Steenbrugge2005a};
(Ste05b) \citet{Steenbrugge2005b};
(Ste09) \citet{Steenbrugge2009};
(Ter09) \citet{Terashima2009};
(Tom10c) \citet{Tombesi2010c};
(Tom10d) \citet{Tombesi2010d};
(Tom11a) \citet{Tombesi2011a};
(Tom11b) \citet{Tombesi2011b};
(Tom12a) \citet{Tombesi2012a};
(Tom12b) \citet{Tombesi2012b};
(Tom13a) \citet{Tombesi2013a};
(Tom13b) \citet{Tombesi2013b};
(Tom14a) \citet{Tombesi2014a};
(Tom14b) \citet{Tombesi2014b};
(Tom15) \citet{Tombesi2015};
(Tom16) \citet{Tombesi2016b};
(Tom17a) \citet{Tombesi2017a};
(Tom17c) \citet{Tombesi2017c};
(Tor10) \citet{Torresi2010};
(Tor12) \citet{Torresi2012};
(Tor17) \citet{Tortosa2017};
(Tor22) \citet{Tortosa2022};
(Tur05) \citet{Turner2005};
(Tur08) \citet{Turner2008};
(Tur18) \citet{Turner2018};
(Urs16) \citet{Ursini2016};
(Urs18) \citet{Ursini2018};
(Urs19) \citet{Ursini2019b};
(Ven18) \citet{Venturi2018};
(Vig15) \citet{Vignali2015};
(Wal18) \citet{Walton2018};
(Wal19) \citet{Walton2019};
(Wal20) \citet{Walton2020b};
(Wal21) \citet{Walton2021};
(Wan05) \citet{Wang2005};
(Win10) \citet{Winter2010c};
(XuY21) \citet{XuYerong2021};
(XuY22) \citet{XuYerong2022};
(XuY23) \citet{XuYerong2023};
(Yaq03a) \citet{Yaqoob2003a};
(Yaq03b) \citet{Yaqoob2003b};
(You05) \citet{Young2005};
(Zha11) \citet{Zhang2011};
(Zhe08) \citet{Zheng2008};
(Zog10) \citet{Zoghbi2010};
(Zog19) \citet{Zoghbi2019}.
\fi
}
\tablenotetext{\ }{(This table is available in its entirety in machine-readable form.)}
\end{deluxetable*}

\clearpage



\bibliography{sample631}{}
\bibliographystyle{aasjournal}



\end{document}